\documentclass{aa}  
\usepackage{graphicx}
\usepackage{txfonts}
\usepackage{booktabs}

\begin{document}

   \title{Temperature constraints from inversions of synthetic solar optical, UV and radio spectra} 

   \subtitle{}

   \author{J. M. da Silva Santos\inst{1} \and J. de la Cruz Rodríguez\inst{1} \and J. Leenaarts\inst{1}
          }
   \institute{Institute for Solar Physics, Department of Astronomy, Stockholm University, AlbaNova University Centre, SE-106 91 Stockholm, Sweden, \email{joao.dasilva@astro.su.se}
             }
   \date{\today}

  \abstract
  % context heading (optional)
   {High-resolution observations of the solar chromosphere at millimeter wavelengths are now possible with the Atacama Large Millimeter Array (ALMA), promising to tackle many open problems in solar physics. Observations from other ground and space-based telescopes will greatly benefit from coordinated endeavors with ALMA, yet the diagnostic potential of combined optical, ultraviolet and mm observations has remained mostly unassessed.} 
  % aims heading (mandatory)
   {In this paper we investigate whether mm-wavelengths could aid current inversion schemes to retrieve a more accurate representation of the temperature structure of the solar atmosphere.}
  % methods heading (mandatory)
   {We performed several non-LTE inversion experiments of the emergent spectra from a snapshot of 3D radiation-MHD simulation. We included common line diagnostics such as \ion{Ca}{ii} H, K, 8542 \AA~and \ion{Mg}{ii} h and k, taking into account partial frequency redistribution effects, along with the continuum around 1.2 mm and 3 mm.}
  % results heading (mandatory)
   {We found that including the mm-continuum in inversions allows a more accurate inference of temperature as function of optical depth. The addition of ALMA bands to other diagnostics should improve the accuracy of the inferred chromospheric temperatures between $\log \tau\sim[-6,-4.5]$ where the \ion{Ca}{ii} and \ion{Mg}{ii} lines are weakly coupled to the local conditions. However, we found that simultaneous multi-atom, non-LTE inversions of optical and UV lines present equally strong constraints in the lower chromosphere and thus are not greatly improved by the 1.2 mm band. Nonetheless, the 3 mm band is still needed to better constrain the mid-upper chromosphere.}
  % conclusions heading (optional), leave it empty if necessary 
   {}


   \maketitle

\section{Introduction}

The solar chromosphere is made of a partially ionized magneto-fluid that is opaque in the core of H$\alpha$, conferring its characteristic fibrilar fine structure on disk and spicular appearance at the limb \citep[e.g.][]{2003ApJ...590..502D, 2004Natur.430..536D, 2007ApJ...655..624D,2013ApJ...776...56R,2017A&A...597A.138R,2017A&A...598A..89R}. This layer is of special interest because it is not clear what are the most significant heating mechanisms among the numerous candidates such as acoustic and MHD waves, magnetic reconnection, microflares and spicules, that can explain the energy balance \citep{1977ARAA..15..363W,1990SSRv...54..377N}. 

In addition to H$\alpha$, one usually resorts to complicated diagnostics formed under non-local thermodynamic equilibrium (non-LTE) conditions such as the \ion{Mg}{ii} h and k and \ion{Ca}{ii} H and K resonance doublet lines, moreover affected by scattering and partial frequency redistribution (PRD). These have been used to observe the chromosphere and infer temperatures and velocities, while the \ion{Ca}{ii} 8542 \AA~infrared line has also been commonly used to probe chromospheric magnetic fields \citep[see review by][]{2017SSRv..210..109D}. 

On the other side of the spectrum, sub-mm/mm continua have the advantage of being formed under LTE in the chromosphere, therefore the source function is Planckian and the intensity is a linear function of the electron temperature since the Rayleigh-Jeans limit applies \citep[e.g][]{2003rtsa.book.....R}. However, LTE does not hold to describe the ionization degree of hydrogen in the chromosphere where time-dependent non-equilibrium effects should be taken into account to compute opacities \citep{2006A&A...460..301L}. These are set by thermal bremsstrahlung processes in the quiet sun \citep{2002AN....323..271B,2017A&A...598A..89R}, namely the electron-proton (H$^{0}$) and electron-neutral (H$^{-}$) free-free absorptions, having a quadratic dependence on wavelength. This in turn means that mm radiation becomes optically thick at increasing heights sampled by increasing wavelengths \citep{2007A&A...471..977W}.

In active regions, a contribution from gyrosynchroton is expected, whereas cyclotron radiation (gyroresonance) would require unrealistically strong magnetic fields with strengths of 28\,000 G and 3300 - 6000 G for this emission to be significant at 1.2 mm and 8 mm respectively \citep{2009A&A...493..613B}. 

Radio observations are thus great alternative chromospheric probes, and can be used to test models of the solar atmosphere \citep{2008Ap&SS.313..197L}. The problem had been that such observations \citep[e.g][]{1992ApJ...384..656W,2006A&A...456..697W, 1993ApJ...418..510B,2014A&A...561A.133L,2015ApJ...804...48I,2016ApJ...816...91I} have had low angular resolution ($\gg$ 5\arcsec) for a long time, which is critical to observe the solar chromosphere that shows rapidly evolving fine structures at least down to the diffraction limit of today's 1m-class optical telescopes.

Although that is promised to change with the advent of the Atacama Large Millimeter Array \citep[ALMA, ][]{2009IEEEP..97.1463W}, providing for the first time a high spatial and temporal resolution view of the solar chromosphere in the radio, interferometric imaging is not a straightforward task. The existence of both extended and localized sources and their variability on very short time scales, and other practical issues, make it difficult to perform image reconstruction of the Sun \citep{2017SoPh..292...87S}. Even if that is achieved, ALMA observations will still benefit from coordinated observations with other ground facilities such as the Swedish Solar Telescope \citep[SST,][]{Scharmer} or GREGOR \citep{2012ASPC..463..365S}, and the upcoming Daniel K. Inouye Solar Telescope (DKIST), and space telescopes like Hinode \citep{2007SoPh..243....3K} or the Interface Region Imaging Spectrograph \citep[IRIS,][]{2014SoPh..289.2733D}.

Spectral lines can provide complementary information on the plasma parameters such as temperature, velocities and magnetic field. The latter is specially important because at least in the first years of ALMA solar observations (Cycles 4, 5 and 6), polarimetry is not commissioned yet. In the future, the study of the continuum circular polarization could also be an alternative method of determining the longitudinal component of the chromospheric magnetic field \citep[e.g.][]{2017A&A...601A..43L}.

Studies of synthetic mm images have been done in the past, but the combined diagnostic potential of ALMA mm-bands with visible and UV wavelengths remains largely unassessed. 

\cite{2007A&A...471..977W} were the first to study sub-mm/mm brightness synthesized from 3D radiation-MHD simulations of the quiet sun using an LTE equation of state (EoS) and radiative transfer treated in LTE and non-LTE. Their brightness maps show filamentary brightenings resulting from shock-induced thermal structure, and fainter regions in between, while the average brightness temperature ($T_{\rm b}$) and relative contrast were found to increase with wavelength.

\cite{2015A&A...575A..15L} obtained $T_{\rm b}$ maps from a more realistic Bifrost simulation by \citet{2016A&A...585A...4C} which includes non-equilibrium hydrogen ionization. The conclusion was that mm-brightness can be a measure of the chromospheric thermal structure at the height at which the radiation is formed. However this height is not unique and unambiguous owing to the complex thermal structure. They found that $T_{\rm b}$ is a reasonable measure (albeit not perfect) of the gas temperature at the effective formation height, defined as the height corresponding to the centroid of the contribution function. Furthermore, their results indicate that although instrumental smearing reduces the correlation between brightness and temperature, the former can still be used to diagnose electron temperatures up to a resolution of 1\arcsec~with reasonable accuracy. Thus spatial resolution becomes a critical issue for observing the solar chromosphere with ALMA.

The first study combining actual solar ALMA data with IRIS data was carried out by \citet{2017ApJ...845L..19B} who found a correlation between $T_{\rm b}(\rm 1.2\,mm)$ and the radiation temperature of \ion{Mg}{ii} h, with a slope smaller than one, highest in plage and lowest in umbra and quiet-Sun regions. This is probably consequence of the local decoupling of \ion{Mg}{ii} due to the scattering component in its source function, although there could also be systematic differences in the formation heights of both diagnostics. In any case this study showed how complementary IRIS and ALMA observations are, and how useful their comparison can be.

Inferring physical properties of the plasma in the solar atmosphere from spectropolarimetric observations involves inverting the radiative transfer equation \citep[RTE,][and references therein]{delToroIniesta2016}. In this paper we used the new STockholm Inversion Code \citep[STiC, ][]{2016ApJ...830L..30D}, which is capable of dealing with simultaneous, LTE/non-LTE, multi-atom modeling of a large number of frequencies, to investigate whether mm bands could help constraining chromospheric temperatures while improving upon common inversion schemes, for example, using the \ion{Fe}{i} 6301, 6302 \AA~Zeeman triplet, \ion{Ca}{ii} 8542 \AA~and H and K or \ion{Mg}{ii} h and k. To this end we performed multiple tests with a 3D radiation-MHD model and different combinations of spectral lines. 
We want to address the question: do radio observations help state-of-the-art inversions of the solar atmosphere?

\section{Model}
\label{Section:Models}

Temperature, velocities and magnetic fields have a clear imprint on the synthesized line strengths and shapes, therefore it is important to test a fairly large range of physical conditions not only to assess the response of the diagnostics to changes in the plasma parameters, but also to investigate where in the atmosphere the combination of multi-wavelength inversions is most useful.

\begin{figure*}
\centering
\includegraphics[width=\linewidth]{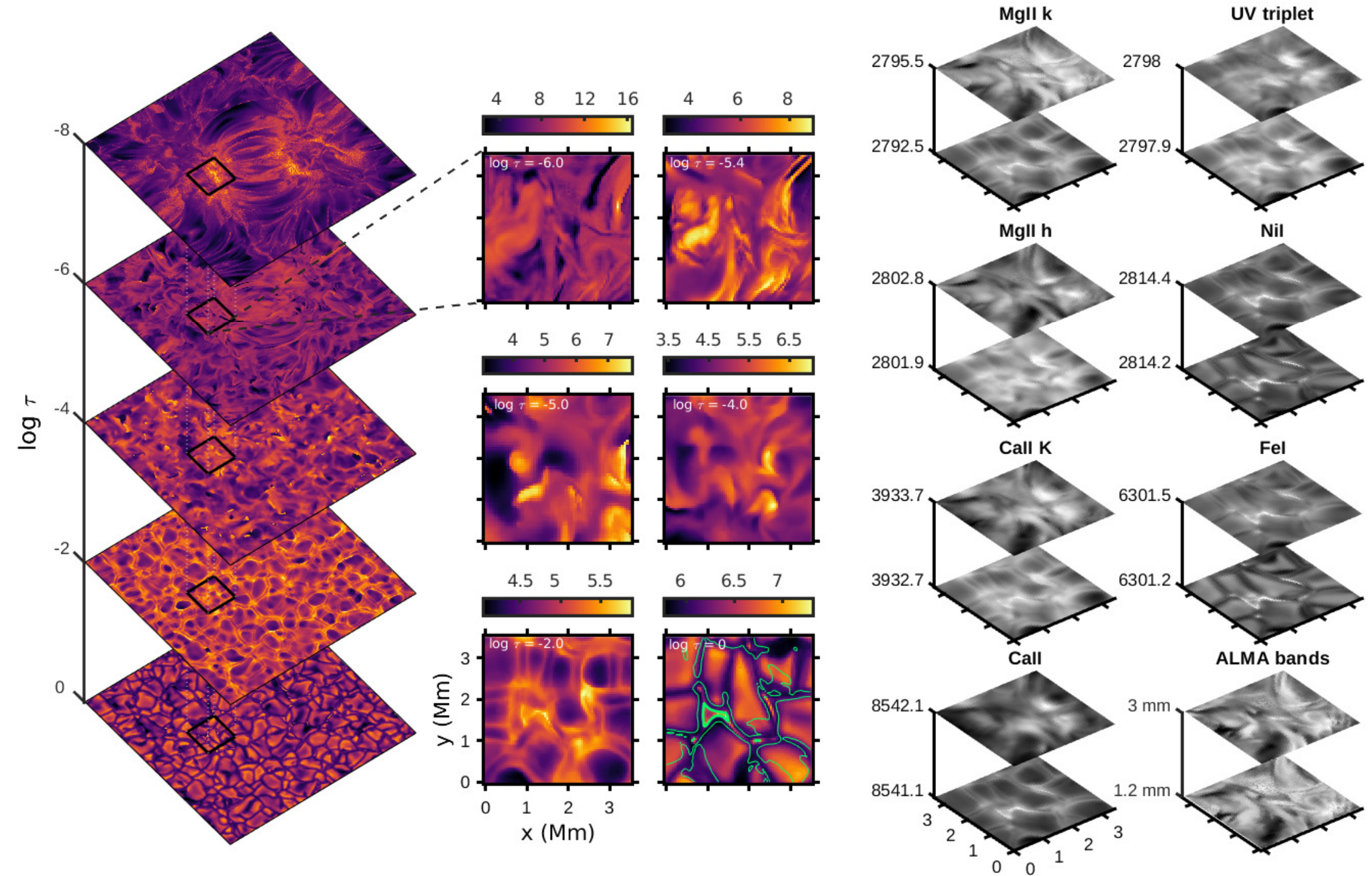} 
\caption{Temperature structure and emerging intensities from a Bifrost simulation. Left panel: the gas temperature as function of optical depth; Central panel: the subregion selected for the inversions with temperature scale in units of $10^3$ K; the magnetic field strength at $\log \tau=0$ is represented by thick (2 kG) and thin (0.1 kG) green isocontours. Right panels: emerging intensities at selected wavelengths given in ångströms except where indicated; for each spectral line, the top and bottom images correspond to core and blue wing wavelengths respectively.}
\label{fig:bifrost}
\end{figure*}

We used snapshot 385 of the radiation-MHD simulation of an enhanced network ("en024048\_hion") performed with the Bifrost code \citep{2011A&A...531A.154G}. This snapshot has been extensively used in the literature to study the formation of chromospheric lines and mm-continua \citep[e.g.][]{2013ApJ...772...89L,2013ApJ...772...90L,2013ApJ...778..143P,2013ApJ...764L..11D,2015ApJ...803...65S,2016ApJ...830L..30D,2016MNRAS.459.3363Q,2017MNRAS.464.4534Q,2015A&A...575A..15L,2017A&A...601A..43L}.

The computational box has $504\times504\times496$ grid points with a pixel scale of approximately 48 km in the horizontal direction, making a domain with 24$\times$24 Mm$^2$, and variable grid separation in the vertical direction from 19 km in the photosphere to 100 km in the corona. 

The simulation includes optically thick radiative transfer in the photosphere and low chromosphere, and parameterized radiative losses in the upper atmosphere \citep{2012A&A...539A..39C}. It takes into account thermal conduction along magnetic field lines and an EoS that includes the effects of non-equilibrium ionization of hydrogen \citep{2007A&A...473..625L}. For additional references and a detailed description of this simulation we refer to \citet{2016A&A...585A...4C}.

However, it lacks non-equilibrium ionization of helium and the effects of partial ionization on chromospheric heating by magnetic fields \citep{2013ApJ...772...90L}. Furthermore, comparisons between synthetic observables and real data by those authors showed that the simulated chromospheric lines tend to be relatively weaker and narrower. Even so, the overall agreement between the simulated features and the real data is good, implying that much of the relevant physics is included. 

\citet{2017A&A...598A..89R} noted that $\rm H\alpha$ fibrils synthesized from this simulation are less numerous and opaque than real solar fibrils, which in turn should be as opaque or even more opaque at mm-wavelengths. This is also not how the synthetic ALMA images obtained by \citet{2015A&A...575A..15L} from the same Bifrost snapshot look like. Therefore one has to keep in mind that the simulation has certain limitations. 

The computational cost of inversions scales with the number of pixels and it is currently prohibitive to perform several cycles of inversions using different combinations of lines for the entire simulation box treating each of the 254\,016 pixels individually since each one can take between $\sim$ 2-6 h to converge depending on the atmosphere, the exact numerical settings and the number of frequency points and depth nodes used. For this reason we restrain the analysis to a more modest box of $74\times74$ pixels (3.5$\times$3.5 Mm$^2$) that encloses one of the magnetic patches and it is large enough for some structure to be recognizable while providing sufficient statistics. 

Since neither the \ion{Ca}{ii} nor the \ion{Mg}{ii} lines are sensitive to temperatures much higher than 16\,000 K \citep{2012A&A...539A..39C}, the coronal-like part was cropped out as well as the domain below $z < -0.2$ Mm, keeping only formation heights of the studied lines. 

The two leftmost panels in Fig. \ref{fig:bifrost} show the temperature stratification in the snapshot as function of optical depth. Throughout this paper we abbreviate the optical depth at 500 nm ($\log\tau_{500}$) simply as $\log\tau$. 
The black square illustrates the selected box for the calculations. At $\log \tau= 0$ we clearly see the granulation in the photosphere with hot granules with temperatures between $T\sim6200-7000$ K, whereas the darker intergranular lanes are cooler ($T\sim5800$ K), and overall the average temperature is 6300 K. Going higher up to $\log \tau=- 2$, which corresponds to $z\sim250-350$ km, the average temperature drops to approximately 4950 K and the reverse granulation becomes visible featuring dark granules with temperatures below 4500 K and brighter intergranular patches with temperatures between ($T\sim5000-6000$ K). In the chromosphere a given optical depth can correspond to very different geometrical heights, depending on the underlying density stratification and, to some extent, the magnetic field from pixel-to-pixel. For example, $\log \tau=- 6$ corresponds to a broad height range usually within $z\sim1300-1800$ km. The gas temperature at this depth shows a filamentary structure connecting the magnetic patches of opposite polarity with cold pockets of only $T\sim2500$ K plasma juxtaposed with hotter structures up to $T\sim11\,700$ K.

For the purpose of comparing response functions only we also extracted a diagonal slice running through the two magnetic patches defined by [$x_0$=50, $y_0$=300, $x_1$=450, $y_1$=150] in pixel coordinates that is approximately 21 Mm in physical length in the simulation snapshot (Section \ref{sec:RF:BIF}).

\begin{table*}
\centering
\caption{List of inversion schemes.}
\label{tab:schemes}
\begin{tabular}{llc}
\hline\hline
\multicolumn{2}{l}{Inversion schemes} & Data \\ \hline
i1 & IRIS & \ion{Mg}{ii} h,k, UV triplet \\
i2 & IRIS+ALMA & \ion{Mg}{ii} h, k, UV triplet; \ion{Ni}{i} 2814.4 \AA; ALMA 1.2 and 3 mm  \\
i3 & SST/CRISP & \ion{Fe}{i} 6301, 6302 \AA; \ion{Ca}{ii} 8542 \AA \\
i4 & SST/CRISP+ALMA & \begin{tabular}[c]{@{}c@{}}\ion{Fe}{i} 6301, 6302 \AA; \ion{Ca}{ii} 8542 \AA; ALMA 1.2 and 3 mm\end{tabular} \\
i5 & IRIS+SST & \begin{tabular}[c]{@{}c@{}}\ion{Fe}{i} 6301, 6302 \AA; \ion{Ca}{ii} 8542 \AA, H, K; \ion{Mg}{ii} h, k, UV triplet\end{tabular} \\
i6 & IRIS+SST+ALMA & \begin{tabular}[c]{@{}c@{}}\ion{Fe}{i} 6301, 6302 \AA; \ion{Ca}{ii} 8542 \AA, H, K; \ion{Mg}{ii} h, k, UV triplet; ALMA 1.2 and 3 mm\end{tabular} \\ \hline
i7 & IRIS+ALMA & \begin{tabular}[c]{@{}c@{}}\ion{Mg}{ii} h, k, UV triplet, \ion{Ni}{i} 2814.4 \AA, ALMA 1.2 mm\end{tabular} \\ \hline
\end{tabular}
\end{table*}

\section{Calculations}
\label{Section:Calculations}

We performed the calculations using the STiC code \citep{2016ApJ...830L..30D}, which is based on the RH code \citep{2001ApJ...557..389U}, to synthesize and invert the spectral data. STiC is capable of simultaneously inverting the Stokes profiles of multiple LTE/non-LTE lines (and the mm-continuum) including PRD effects using node parametrization. It uses third-order DELO-Bezier splines to integrate the RTE \citep{2013ApJ...764...33D} and it assumes plane-parallel geometry and hydrostatic equilibrium. Further details and more recent updates on the code can be found in \textbf{de la Cruz Rodriguez et al (in prep.)}.

\subsection{Forward calculations}
\label{Section:Calculations:Forward}
The spectral lines and the radio continuum were synthesized column-by-column (1.5D approximation) using an LTE EoS while the statistical equilibrium is solved for non-LTE populations of a five-level plus continuum hydrogen atom model, a five-level plus continuum model for \ion{Ca}{ii} with PRD in the H and K lines and complete frequency redistribution (CRD) in the 8542 \AA~line, and ten-level plus continuum model for \ion{Mg}{ii} with PRD in the h and k lines and CRD in the subordinate UV triplet. For \ion{Fe}{I} we used a fifteen-level plus continuum model assuming LTE and CRD.  
The radiation is assumed to come from the solar disk center.

In the solar chromosphere full 3D radiative transfer is important for strongly scattering lines \citep{2009ApJ...694L.128L,2012ApJ...749..136L,2013ApJ...772...89L}. Moreover, PRD effects have to be taken into account. This has been done in full 3D for the first time in \citet{2017A&A...597A..46S} for \ion{Mg}{ii} h and k, and in \citet{2017arXiv171201045B} for \ion{Ca}{ii} H and K based on the same Bifrost simulation we use in this paper. This sophisticated treatment comes with an enormous increase in computational cost, and remains prohibitive in inversion codes. In rigor such computations would mostly be necessary in the cores of \ion{Mg}{ii} h and k between the emission peaks \citep{2013ApJ...772...90L}, which we neglect here. In practice this results in \ion{Mg}{ii} intensity maps of higher contrast than reality at core wavelengths. This is also true for the analogous \ion{Ca}{ii} resonance lines. For the same reason we do not include H$\alpha$, the classic chromospheric line, because a 3D treatment of the RTE is crucial due to its highly scattering nature \citep{2012ApJ...749..136L}. This is not yet possible with current inversion codes. Finally, these effects are not as strong in the \ion{Ca}{ii} 8542 \AA~\citep{2012A&A...543A..34D} because it has a deeper formation height, so it can be modeled in 1.5D to good approximation. 

We also included in the synthesis a weak photospheric line of \ion{Ni}{i} 2814.4 \AA~in the IRIS spectral window which, according to \citet{2016ApJ...830L..30D}, helps to infer vertical velocities in the photosphere.

Note that this study is not exhaustive in finding the best lines to invert together with ALMA data. Rather, it concerns the most common diagnostics, and in particular those that can be observed by IRIS and SST, and how ALMA helps inverting those. Other combinations of lines that could have been interesting to investigate include \ion{Mg}{i} $\rm b_2$ 5173 \AA~and \ion{Na}{i} $\rm D_1$ 5896 \AA, forming in the upper photosphere \citep{1995A&A...299..596B, 2010ApJ...709.1362L}, or even \ion{Ca}{ii} 8498 \AA~which is also temperature-sensitive in the chromosphere \citep[e.g.][]{2017MNRAS.464.4534Q}.

We did not take into account gyrosynchroton emission at mm-wavelengths since it would only originate from very energetic positrons and electrons primarily at flaring sites \citep{2016SSRv..200....1W}, which is not the scope of the model atmospheres we were interested in.

The polarization is assumed to come solely from the Zeeman effect, which means that we do not include the circular polarization of the mm-continuum in the calculations, so we cannot investigate in this paper whether ALMA helps constraining magnetic fields. 

In real observations it is customary to introduce an additional parameter to explain the observed line broadening that does not correspond to thermal or instrumental origin, but could be due to a non-thermal contribution or simply unresolved motions in the same pixel. This micro-turbulence broadening is assumed to have a gaussian line-of-sight velocity distribution such that the total Doppler broadening $v_{\rm D}^2 = \sqrt[]{v_{\rm th}^2+v_{\rm turb}^2}$ is a sum in quadrature of the thermal ($v_{\rm th}^2=2kT/m$) and turbulent components.
We investigated two cases regarding the micro-turbulence velocity ($v_{\rm turb}$): (1) a constant $v_{\rm turb}(\log \tau)=$ 2 $\rm km~s^{-1}$ which has a line broadening effect, and (2) $v_{\rm turb}(\log \tau)=0$.

The near-UV lines were convolved with an instrumental profile with full-width-at-half-maximum (FWHM) of 0.053 \AA~(to simulate IRIS observations), and the optical lines between 0.052-0.1 \AA~(SST/CHROMIS and CRISP). In the mm-continuum we used 8 points centered on the 4 sub-bands of each currently-available ALMA band: [1.20, 1.21, 1.29, 1.30] mm and [2.80, 2.86, 3.16, 3.22] mm. 

We did not add random noise to the data to ensure that the results do not depend on an arbitrary noise level, and we did not take into account the issue of different spatial resolutions, i.e. we assume each telescope perfectly resolves horizontal structures. 

\subsection{Inverse calculations}
\label{Section:Calculations:Inverse}

Having computed the line profiles and mm-continua, we performed several inversion experiments using different combinations of spectral data to evaluate whether ALMA helps to constrain the inversion of non-LTE lines. 

We have tested several inversion schemes using different combinations of lines plus mm-continuum. For clarity we shall call them i1 (IRIS), i2+i7 (IRIS+ALMA), i3 (SST/CRISP), i4 (SST/CRISP+ALMA), i5 (SST+IRIS) and i6 (SST+IRIS+ALMA) as summarized in Table \ref{tab:schemes}. The reason behind this separation is twofold: firstly we want to be able to discern the constraining power of ALMA apart from the lines, so we need to test inversion schemes with and without chromospheric lines, and secondly, such co-temporal/spatial data spanning the entire electromagnetic spectrum is not often available for a given solar feature of interest, but we still want to know to what extent does ALMA help constraining whatever common line observations one disposes, in particular when observations from other ground telescopes are not possible, whereas IRIS from space might still be able to record data.

Although we synthesized the full Stokes vector in each grid point, we fitted only Stokes I because in this paper we are primarily interested in the temperature stratification and not in magnetic fields. Therefore we run the inversions in the "NO\_STOKES" mode. This makes the computations faster and less prone to failed convergence. Because this is a quiet-Sun simulation, the response of the spectral lines to the magnetic field is mostly negligible, so this approximation does not critically affect the conclusions.

The fitting weights were set in a arbitrary absolute scale which means the value of the merit function $(\chi^2)$ is meaningless, but we assigned relatively more weight to the line cores compared to the optical and UV continuum. The mm-wavelength points were assigned even stronger weights (8 times higher than the line cores) to make sure the code always fits the mm-continuum.

We run the inversions in two initial cycles with an increasing number of nodes to improve convergence, and with regularization switched off. Bad pixels are eliminated and replaced with interpolated values at all optical depths using a 2D Gaussian convolution kernel.
Then, we performed a third cycle with regularization. At all cycles the best-fit in each pixel was chosen by selecting the model with the lowest $\chi^2$ among the results of a few randomizations of the initial guess. 

We extensively tested the dependency of the results on the placement and number of nodes, but it was impossible to find an optimal setup that works for any pixel in the atmosphere. On the other hand, it is necessary to minimize the differences between the numerical settings of each set of inversions to ensure that the results are dominated by physics, not by the inversion method. 
 
This being said, we started with [6, 3, 1] nodes in [$T$, $v_{\rm los}$, $v_{\rm turb}$] respectively, and we increased to [10, 6, 1] in cycle 2 and up to [14, 8, 1] in cycle 3 adding a modified Tikhonov regularization \textbf{de la Cruz Rodríguez et al (in prep).} in $T$ and $v_{\rm los}$. The first cycle was run with equidistant node placement, but the two following cycles were run with a more sparse placement at higher optical depths and a more dense sampling at lower optical depths to ensure that the accuracy of the inverted chromosphere is not limited by the number of nodes.

We also performed one additional test running with 0 nodes in $v_{\rm turb}$ on data synthesized with $v_{\rm turb}=0$.

\subsection{Response functions}
\label{Section:rfs}
To better understand the results of the inversions it is instructive to compute response functions ($R_\lambda$) as function of wavelength and optical depth since they reflect the different sensitivities of the diagnostics to changes in the plasma parameters \citep{1975SoPh...43..289B,1977A&A....56..111L}, in particular to temperature as we are interested here.
We numerically computed $R_\lambda$ to temperature by introducing a small first-order perturbation in the atmosphere and comparing the resulting spectra with the pre-obtained. This can be expressed by the following formula:
\begin{equation}
R_{\lambda}(T, \tau_{\rm k}) = \frac{\partial I_\lambda}{\partial T}\bigg\rvert_{\tau_k} \approx \frac{I_{\lambda}(T+\Delta T, \tau_{\rm k})-I_\lambda(T,\tau_{\rm k})}{\Delta T}
\end{equation}
\noindent where $I_\lambda$ is the intensity at a given wavelength and $\Delta T = 10$ K is the magnitude of the temperature perturbation.

Note that the exact shape of $R_\lambda$ as function of wavelength and optical depth is obviously not only model dependent, but also dependent on the magnitude of the perturbation to some extent. The perturbation should be kept small, but large enough to beat the numerical noise \citep{2017A&A...601A.100M}. Numerical tests have shown that the value adopted is robust in the sense that it ensures the linear regime.

\section{Results}
\subsection{Emergent spectra}

\begin{figure*}
\centering
\includegraphics[width=0.97\textwidth]{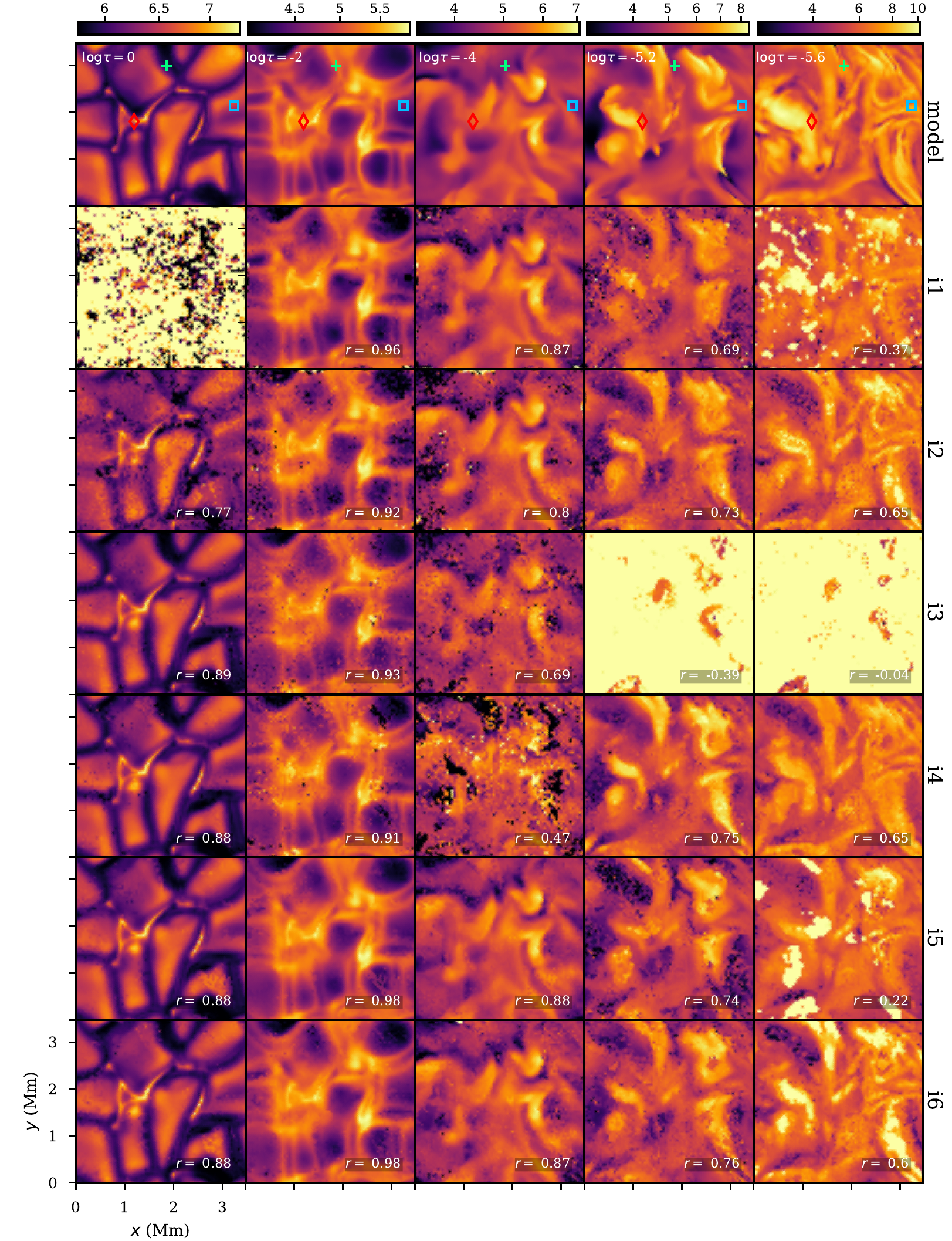} 
\caption{Inverted temperatures compared to the model. Each column corresponds to a given optical depth and each row to a different inversion scheme (not smoothed) which is to be compared to the model in the top row. The colorbars are in units of $10^3$ K. The three markers in the top row correspond to the spectra in Fig. \ref{fig:BIF_specs}. $r$ is the Pearson's correlation coefficient.} \label{fig:BIF_mosaic}
\end{figure*}

\begin{figure}
\centering
\includegraphics[width=\linewidth]{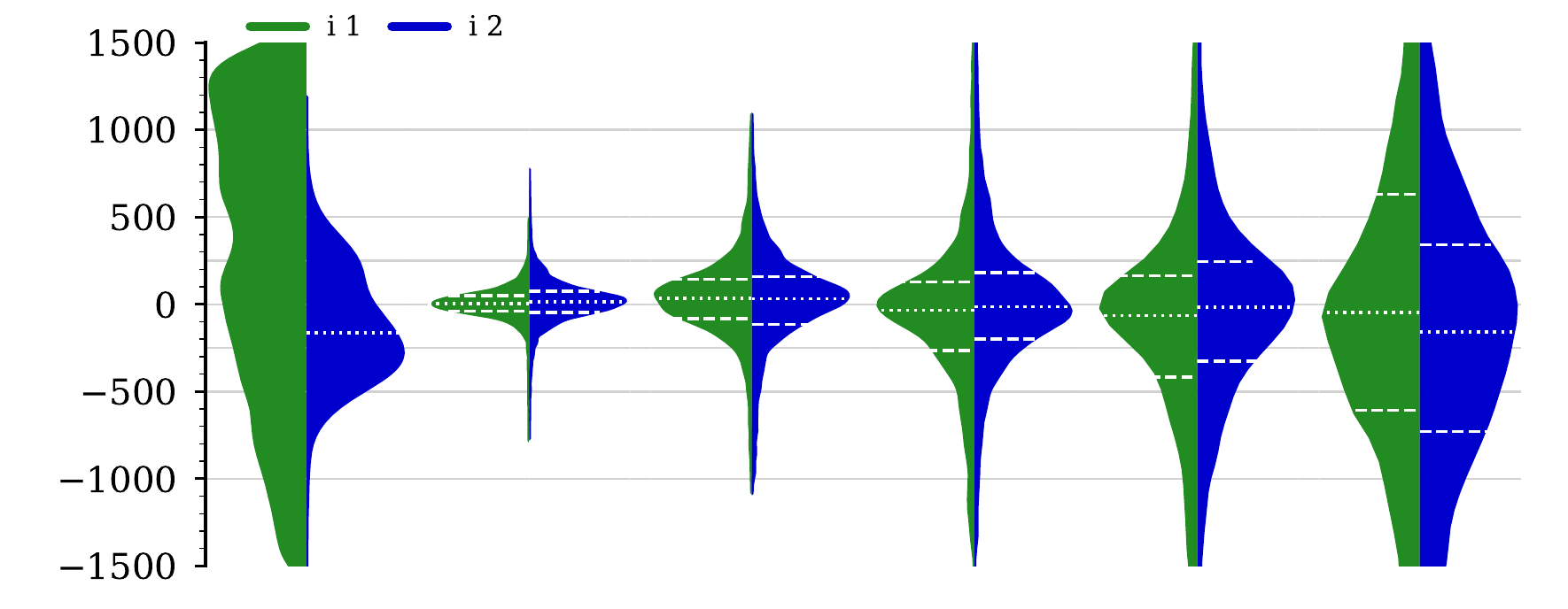}\\
\includegraphics[width=\linewidth]{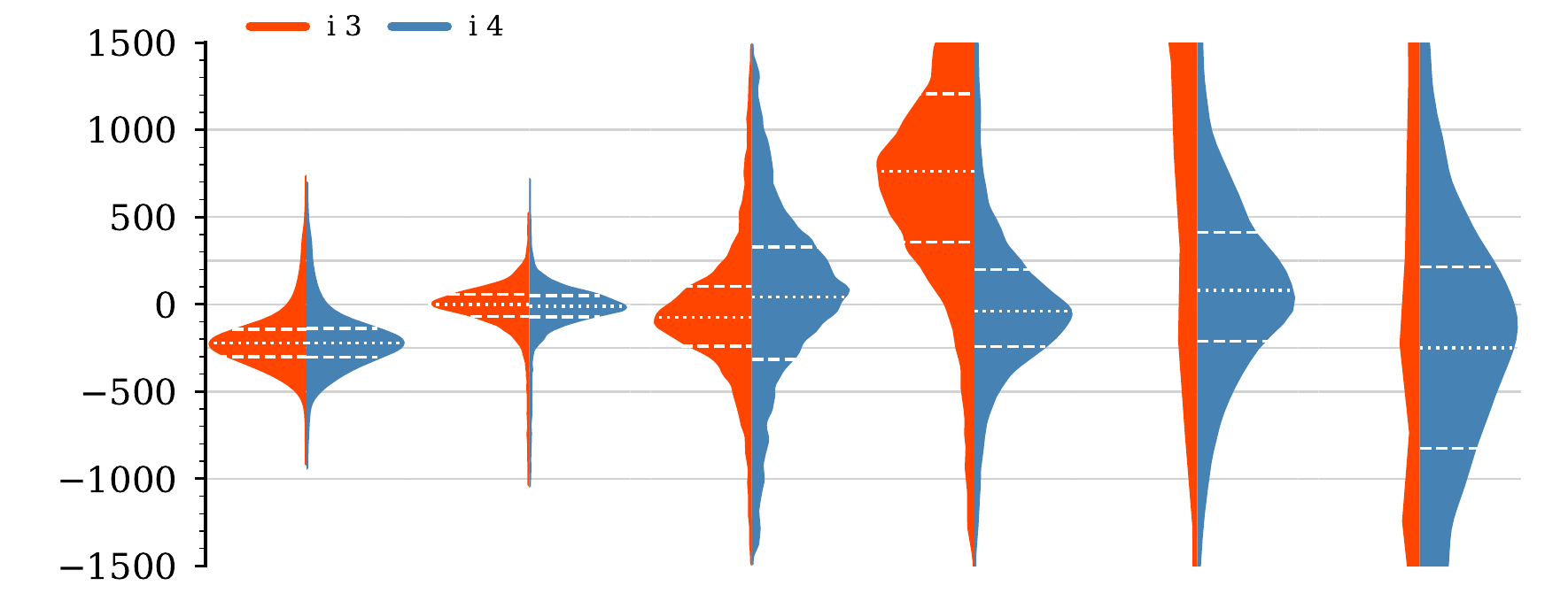}\\
\includegraphics[width=\linewidth]{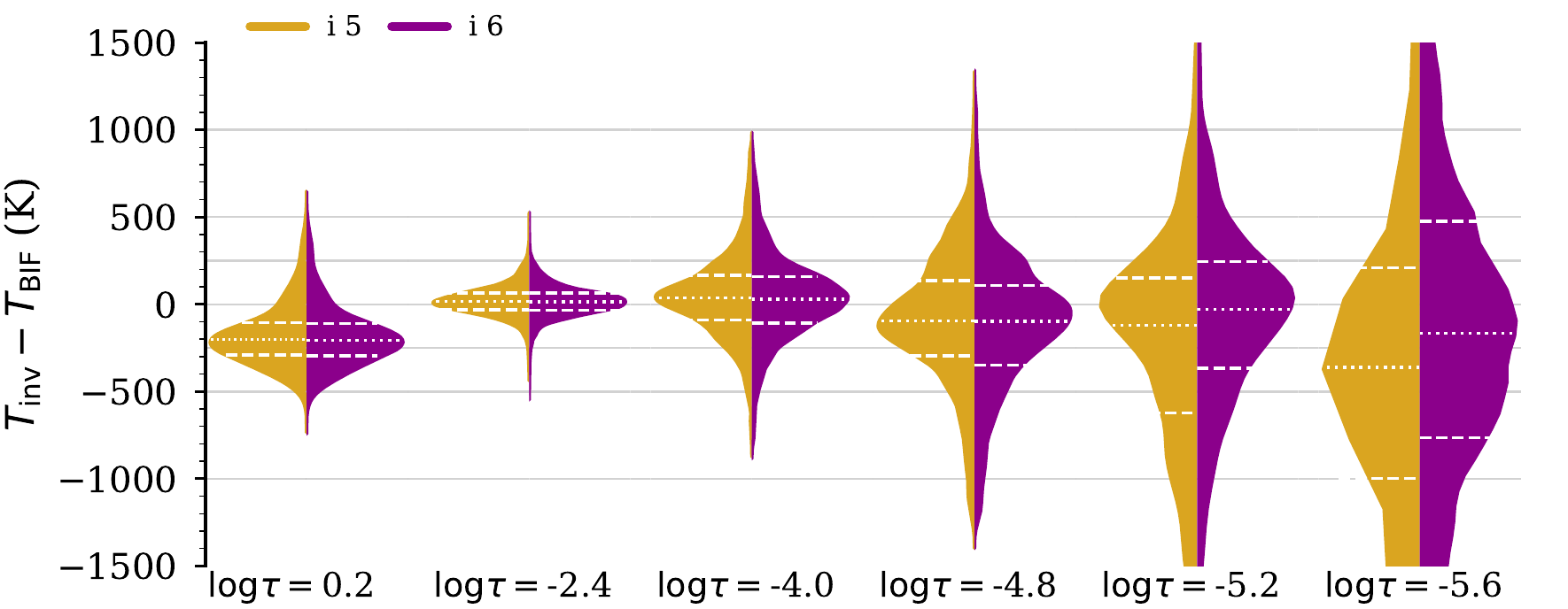}
\caption{Temperature residuals at selected optical depth nodes. Each panel compares the probability density distributions of the temperature residuals of two inversion schemes (with and without mm-continuum), with the median indicated by the white dotted lines while the $16^{\rm th}$ and $84^{\rm th}$ percentiles are shown by dashed lines.} \label{fig:residuals_vl}
\end{figure}

The emerging intensities from the model resemble the temperature maps at a given height (see Fig. \ref{fig:bifrost}). For example, the granulation in the photosphere is very clearly seen at wing-wavelengths of \ion{Ni}{i} 2814.4 \AA ~and \ion{Fe}{i} 6301.5 \AA. The layers above can be probed for example with \ion{Ca}{ii} 8542 \AA~where we see the reversed granulation at $\lambda_{0}-1$ \AA, but also if one scans sufficiently far way from the core of \ion{Ca}{ii} H and K and \ion{Mg}{ii} h and k. The images in the cores of the K and k lines look very similar, showing a somewhat higher layer of the atmosphere, and correlate with the temperature maps between $\log \tau \sim[-6.0, -5.0]$. The formation heights of the continuum at 1.2 mm and 3 mm are known to be similar to the cores of \ion{Mg}{ii} h and k \citep[e.g.][]{2017ApJ...845L..19B}, and in fact one can see that the intensity maps show analogous structures. 

Interestingly we also found a correlation between $T_{\rm b}(\rm 1.2\,mm)$ and the radiation temperature of the core of \ion{Mg}{ii} h with a slope smaller than one as in \citet{2017ApJ...845L..19B} from actual observations. Here we found Pearson's correlation coefficients between 1.2 mm and \ion{Mg}{ii} temperatures approximately equal to 0.78 for the k line and 0.79 for the h line. 

The average spectra from this simulation patch shows \ion{Ca}{ii} H, K and 8542 \AA~in and UV triplet in absorption, whereas the h and k lines of \ion{Mg}{ii} are in emission with double peaks. The mean brightness temperatures (with standard deviation) in the radio are $T_{\rm b}(\rm 1.2\,mm)=$ 5800 ($\pm 880$) K and $T_{\rm b}(\rm 3\, mm)=$ 7250 ($\pm 1140$) K. These values are consistent with recent ALMA single-dish mapping observations. \citet{2017SoPh..292...88W} reported $T_{\rm b}(\rm 1.3\,mm)=$ 5900 ($\pm100$) K and $T_{\rm b}(\rm 3\,mm)=$ 7300 ($\pm100$)K, while \citet{2017arXiv170509008A} obtained $T_{\rm b}(\rm 1.3\,mm)=$ 6180 ($\pm100$) K and $T_{\rm b}(\rm 3\,mm)=$ 7250 ($\pm170$) K. 

\subsection{Inverted temperatures}
\label{sec:BIF_results}

Figure \ref{fig:BIF_mosaic} shows the results of six inversion schemes at five selected optical depths at which we compared model temperature versus inverted temperature. These maps correspond to the final inversion results and were not smoothed to remove bad pixels. We show the Pearson's correlation coefficient as a rough measure of association between $T_{\rm model}$ and $T_{\rm inv}$. This figure is complemented by Fig. \ref{fig:residuals_vl} which illustrates the kernel density distributions\footnote{Computed using the Scikit-learn library \citep{scikit-learn}} of temperature residuals as function of optical depth. 

Overall, we found major performance differences most notably in the chromosphere. In qualitative terms we found: \textbf{1)} a poor temperature reconstruction in the low photosphere in i1 and slightly better in i2, but very good in the other schemes; \textbf{2)} a better inversion of the chromosphere when mm-bands are included in i2, i4 and i6; \textbf{3)} a generally better inversion at all heights when more spectral data is used together, but \textbf{4)} no substantial improvement in the low chromosphere when mm-wavelengths are added to schemes already solving for \ion{Mg}{ii} h and k and \ion{Ca}{ii} H, K and 8542 \AA~simultaneously. The first point was expected since i1 does not contain any information of the deeper layers of the atmospheres, whereas i3-6 contain the photospheric \ion{Fe}{i} lines. Note, however, how the temperature map at $\log \tau=-5.6$ of i2 shows much more structure compared to i1 which tends to overestimate temperatures in this layer. The choice of this particular optical depth is not fortuitous. Rather, it corresponds to the average optical depth of the maximum response to temperature of the 3 mm band as we will show in Section \ref{sec:RF:BIF}. 

The difference between i3 and i4 in this layer is even more striking. 
The temperature in the higher chromosphere above $\log\tau\sim-5.0$ is mostly unconstrained when we use only \ion{Ca}{ii} 8542 \AA, but when mm bands are added we suddenly gain access to information that is critical to reconstruct the stratification at these heights. However, at $\log \tau=-4.0$ we see that the i4 map is a lot more noisy than i3 which is a consequence of difficulties in simultaneously fitting both mm-bands and the wings of \ion{Ca}{ii} 8542 \AA. 

We obtained generally better inversions at all optical depths when we combined visible and UV lines as depicted in the i5 row. The region spanning from the middle of the photosphere to the low chromosphere is more accurately reconstructed than in any of the other schemes, which is shown by the much smoother and noise-free temperature maps at, for example, $\log \tau=-2.0 \rm, -4.0, -5.2$. Higher up in the chromosphere at $\log \tau=-5.6$ we also see that the map shows more structure compared to i1 and i3, and it has cleaned up most of the "inversion noise" seen in i1. This means that the \ion{Ca}{ii} H and K lines offer important constraints in these layers. However, there are still some patches of systematically overestimated temperatures (by more than 1500 K). These are greatly reduced in i6 with the addition of mm bands, but not always. 

\subsection{Error analysis}

\begin{figure}[t]
\includegraphics[width=\linewidth]{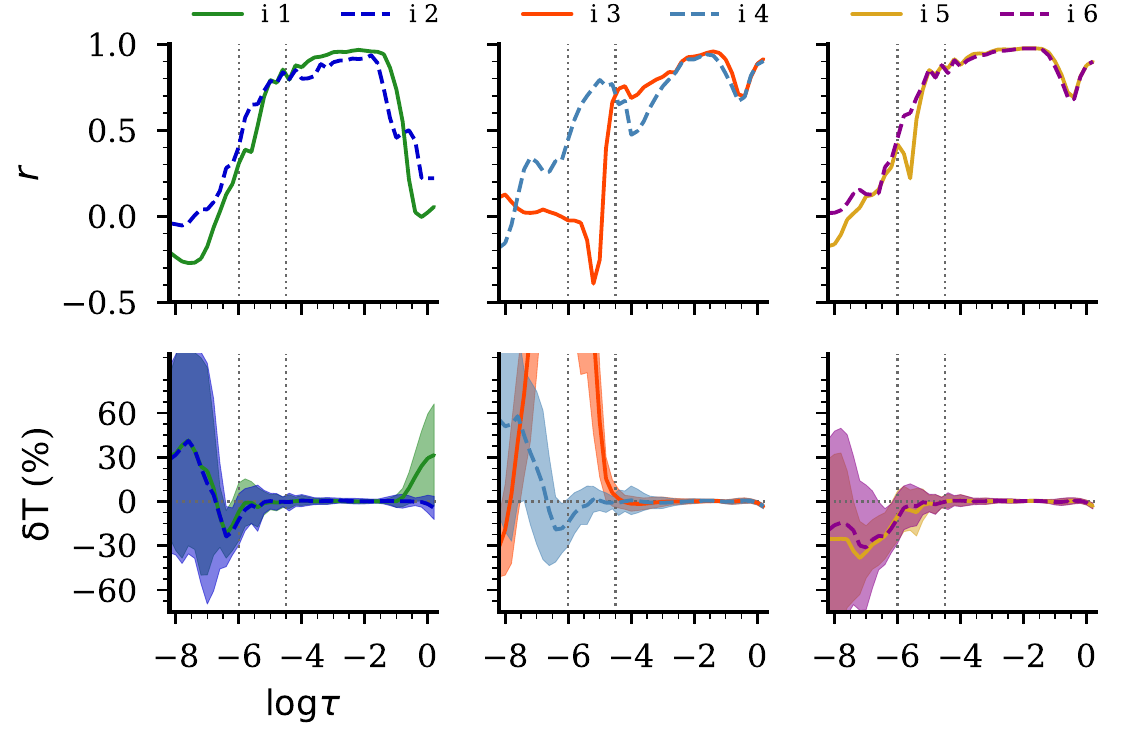}
\caption{Temperature error of the BIF-inversions and product moment correlation coefficient as a function of optical depth.} \label{fig:error}
\end{figure}

We now define the relative temperature error ($\delta T$) and the absolute error ($\Theta$) of the inversion as:
\begin{eqnarray}
\delta T(\log \tau) = \frac{T_{\rm inv}(\log \tau)-T_{\rm model}(\log \tau) }{T_{\rm model}(\log \tau)} \label{eq:error} \\
\Theta(\log \tau) = \lvert T_{\rm inv}(\log \tau)-T_{\rm model}(\log \tau)\rvert \label{eq:error2}
\end{eqnarray}
\noindent where $T_{\rm model}$ is the model temperature at a given optical depth.

Figure \ref{fig:error} shows more quantitative estimates of the error in the inversions. In the top panel we plotted the Pearson's correlation coefficient\footnote{Note that $r$ is aspatial in the sense that it disregards the spatial distribution of the pair of values and assumes that they are independent and non-autocorrelated, which is not true. As a consequence, normal tests of significance cannot be performed. In spite of this, $r$ is still a useful and simple measure of association.} between $T_{\rm BIF}$ and $T_{\rm inv}$ as a function of optical depth. Since $r$ is very affected by strong outliers we made a $4\sigma$-clipping on $T_{\rm inv}$ \textit{a priori}. The exact value of $r$ is not so important, but rather how it evolves with height. Qualitatively this behavior is what one would expect after looking at the results in Fig. \ref{fig:BIF_mosaic}. We found that $r$ tends to decrease with increasing height. These diagrams are specially useful in revealing the loss of correlation for optical depths below $\log \tau \sim -4.5$ which is more severe in the model i1 and i3, but more moderate in the other schemes. We found that schemes with mm-wavelength points maintain a higher correlation with increasing heights compared to the ones without them. However, in the upper photosphere this behavior reverses since the inverted maps at those heights are typically more noisy (see i2 and i4 in  Fig. \ref{fig:BIF_mosaic}). 

In the bottom panel of Fig. \ref{fig:error} we show $\delta T$ as a function of optical depth.
The solid and dashed lines are the median of $\delta T$ for schemes with and without ALMA respectively, whereas the filled contour is a robust measure of its spread given by $\rm \sigma_G=IQR / 2\sqrt[]{2}erf^{-1}(0.5)$ where IQR is the interquartile range and erf is the Gauss error function. The dotted vertical lines delimit a the range of optical depths that contains 95\% of the response functions at 1.2 mm and 3 mm bands (see Fig. \ref{fig:BIF_responses}, discussed further on). 

The plots confirm that the error of the inversion is usually smaller in the photosphere and centered around zero, and larger in the chromosphere where it begins to show a systematic deviation with $T_{\rm inv}$ being more often underestimated. We found that the median absolute error between $\log \tau = [-4,-2]$ is approximately 1-2\% in all schemes. In the photosphere at $\log \tau > -1$ i1 and i2 have a much larger error (>50\%) as expected, but the remaining schemes have $\delta T \la 3$\%. Between $\log \tau \sim [-6,-4.5]$ is where we found the most interesting differences in the error distributions. For example, comparing i1 to i2, we see that the median $\delta T$ oscillates around zero but with a lower dispersion in i2 than in i1 with an IQR that is almost twice as narrow. In this range, i2 has a consistently smaller median absolute error than i1 by 1-5\%. Below $\log \tau=-6$, $\delta T$ greatly increases in both schemes and the $T_{\rm inv}$ are not trustworthy. The difference in $\delta T$ between i3 and i4 in the chromosphere is remarkable. While in i3 the median absolute error peaks at $\sim$250\%, in i4 $\delta T$ it is typically $\la15$\%. As we had seen in Fig. \ref{fig:BIF_mosaic} this is a major improvement. This is not as clear if we compare i5 to i6 because the dispersion in $\delta T$ is similar ($\la10$\%). Between $\log \tau \sim [-6,-4]$ the difference of median absolute errors of i5 and i6 is less than 0.5\%.

These numbers do not tell the complete picture of the error characterization and inversion performance of the schemes since they are robust measures of scale and location which are blind to the tails of the distributions. In fact, despite the difference in the median errors in the chromosphere being arguably small (except for i3/i4, see Fig. \ref{fig:residuals_vl}), the addition of mm-wavelengths is very useful to remove the outliers of the error distributions and substantially reduce the "inversion noise". 

\begin{figure}[t]
\centering
\includegraphics[width=0.99\linewidth]{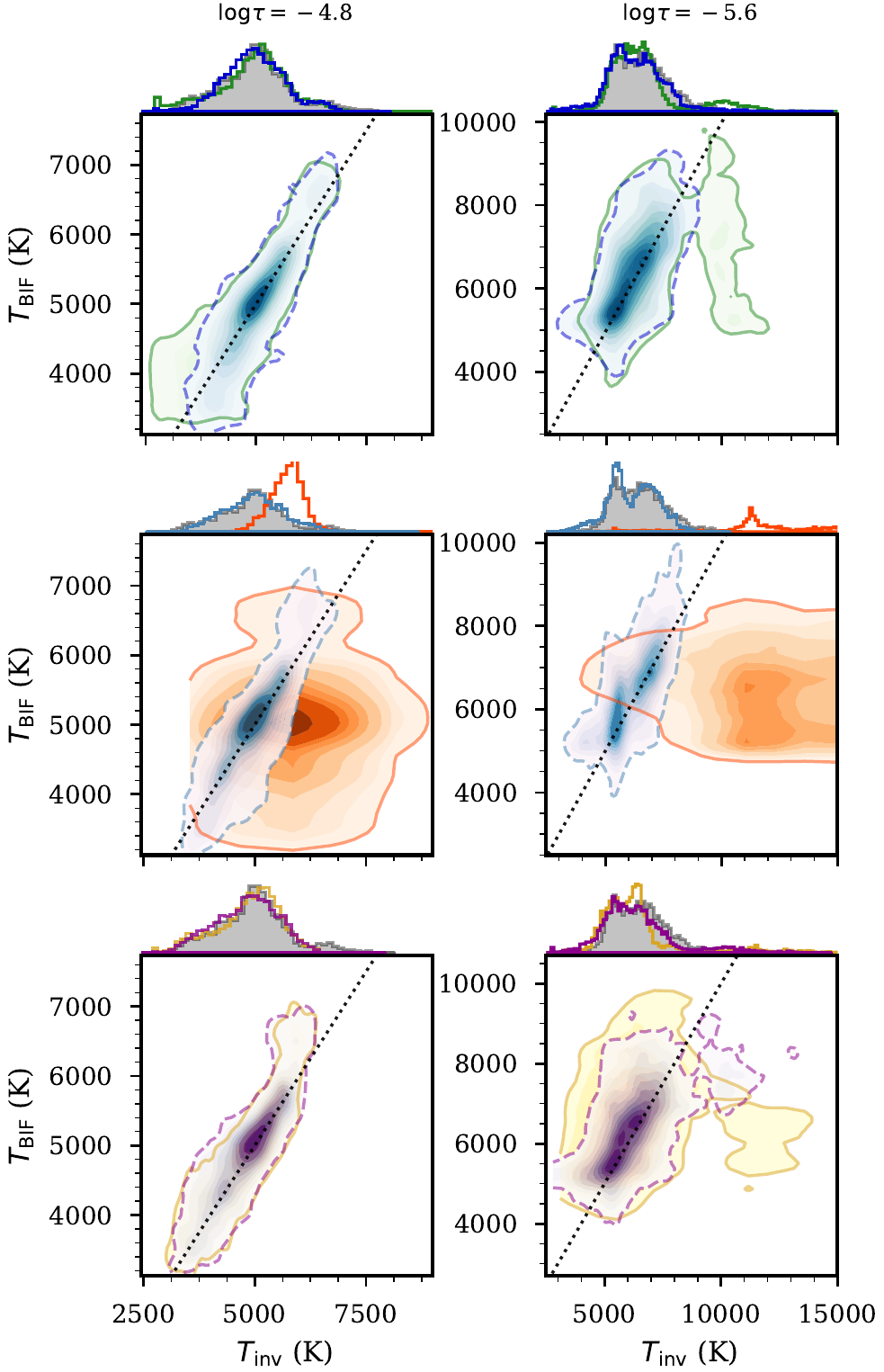}
\caption{Joint probability densities of the BIF model and inverted temperatures at two different optical depths. Each panel compares two inversion schemes (color code as in Fig.\ref{fig:error}), and the dashed and solid lines enclose 90\% of the distributions. The side plots show the corresponding histograms: top row: i1 vs i2, middle: i3 vs i4, and bottom: i5 vs i6. The black dotted lines are the $T_{\rm BIF}=T_{\rm inv}$ locus. } \label{fig:densities}
\end{figure}

Figure \ref{fig:densities} shows the joint probability densities of temperature for the six inversion schemes at two selected optical depths in the chromosphere. On the top side of each panel we show the respective histograms. This figure emphasizes that the differences accentuate with decreasing optical depths, while the dispersion in $T_{\rm inv}$ is reduced with the inclusion of mm- data in the inversions, specially in i1 and i3. This is not only shown by the more compact contours of i2, i4 and i6 around the locus $T_{\rm inv}=T_{\rm BIF}$, but also by their corresponding temperature histograms which, with respect to the histograms of $T_{\rm BIF}$, have smaller relative entropy than i1, i3 and i5. 

\subsection{Examples of fitted spectra}

\begin{figure*}[t]
\centering
\includegraphics[width=0.96\linewidth]{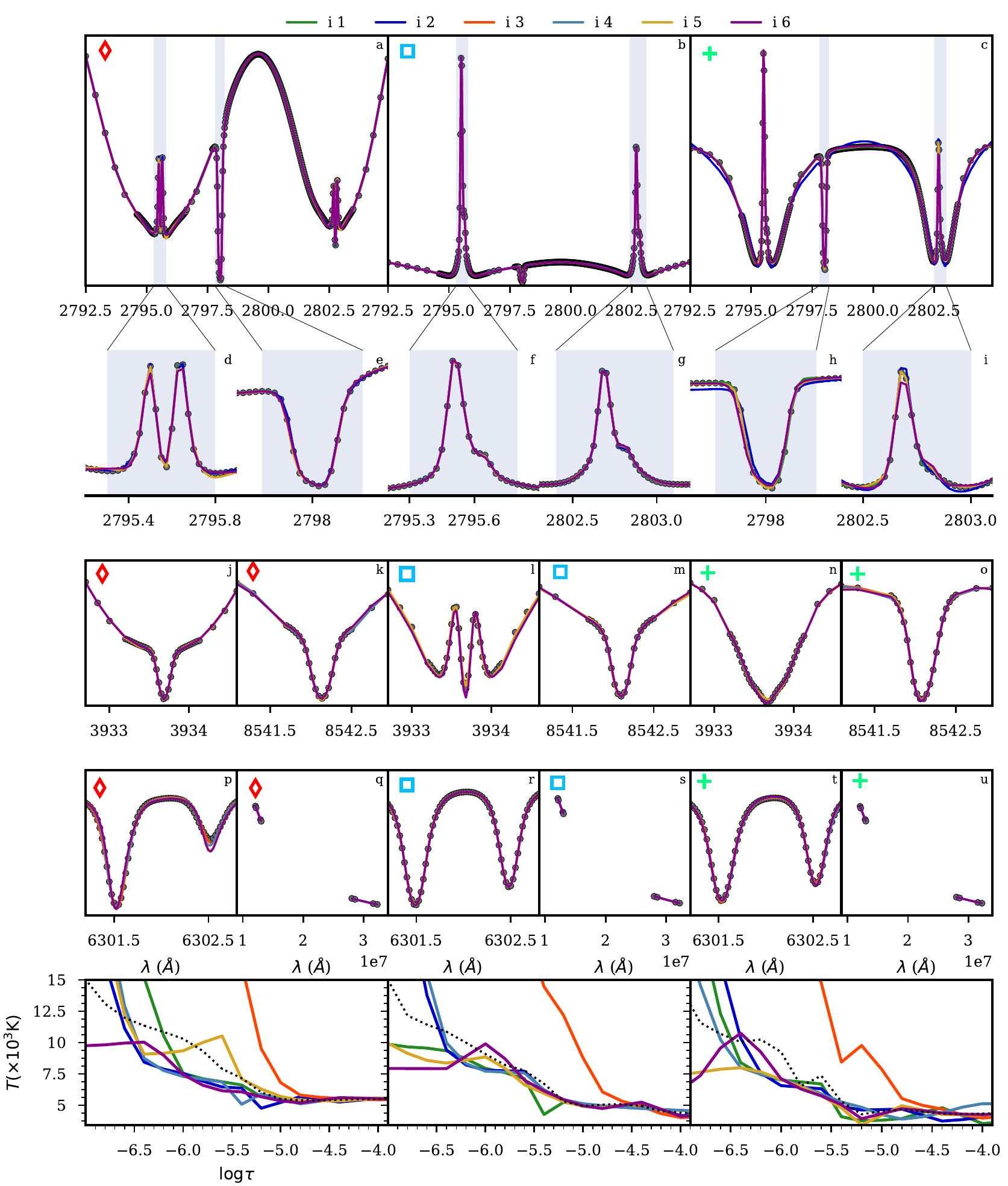}
\caption{Example of fits to the spectra from the Bifrost simulation snapshot. These correspond to the points marked in Fig. \ref{fig:BIF_mosaic}. \textbf{a-c)} overview of \ion{Mg}{ii} h and k, UV triplet lines \textbf{d-i)} a zoom into the cores of \ion{Mg}{ii} h and k, UV triplet; \textbf{j, l, n)} \ion{Ca}{ii} K; \textbf{k, m, o)} \ion{Ca}{ii} 8542 \AA; \textbf{p,r,t)} \ion{Fe}{i} 6301, 6302 \AA; and \textbf{q, s, u)} $\sim$1-3 mm continua. The bottom panels show the model temperature (dotted line) versus the inverted temperatures.} \label{fig:BIF_specs}
\end{figure*}

\begin{figure*}[t]
\centering
\includegraphics[width=\linewidth]{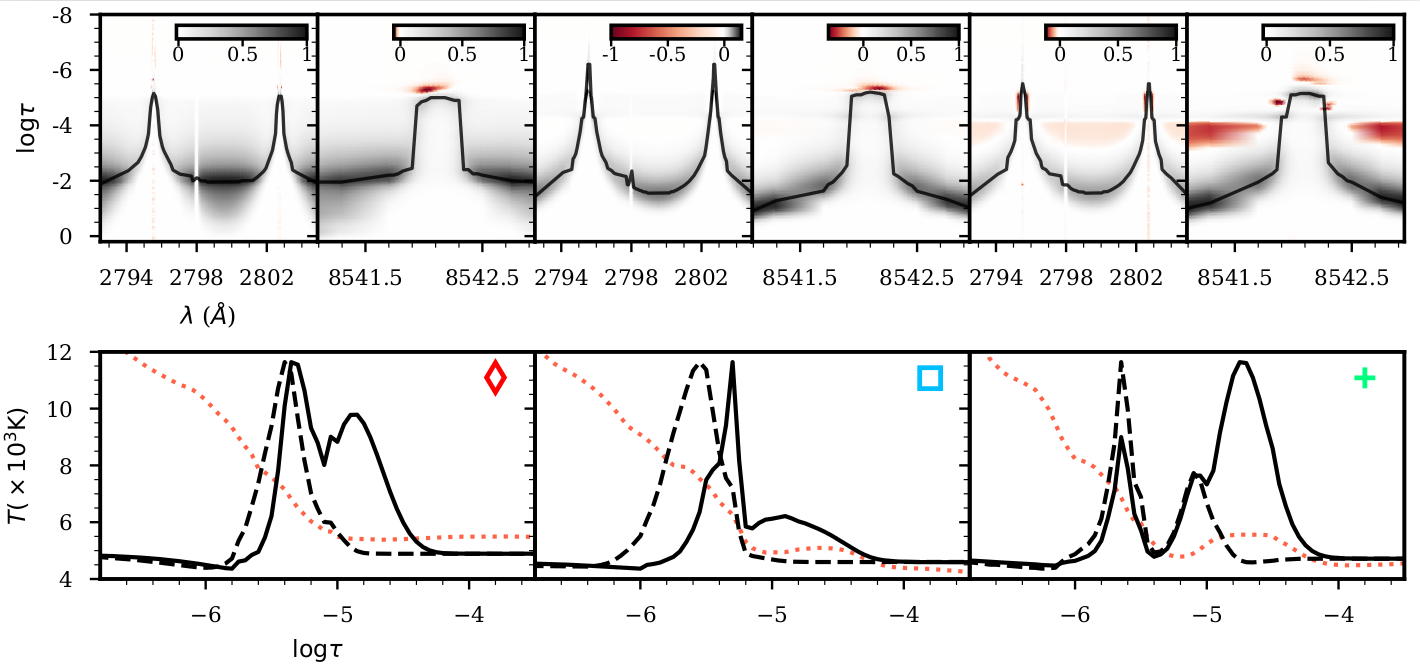}
\caption{Normalized response functions to temperature perturbations at selected frequencies and pixels in the Bifrost model. Top row: response functions to temperature of \ion{Mg}{ii} h and k and \ion{Ca}{ii} 8542 \AA~ at the same locations given in Fig. \ref{fig:BIF_specs}.; the solid line traces the maximum at each column. Bottom row: normalized responses at 1.2 mm (solid line) and 3 mm (dashed line), and model temperature (dotted line) at those locations.} \label{fig:3rfs}
\end{figure*}

Figure \ref{fig:BIF_specs} shows three examples of fitted spectra at the different locations marked in the top row of the mosaic in Fig. \ref{fig:BIF_mosaic}. Overall the spectra are well reproduced with the fit residuals being usually slightly larger in the line cores than in the wings. 

For example, on top of the kilogauss magnetic element (red diamond marker), we obtained very good fits except in the h and k cores. There is also a slight misfit in the cores of the \ion{Fe}{i} lines because we did not include magnetic fields in the inversions, therefore the code cannot correctly reproduce these profiles simply with temperature and velocity changes because they also have a strong response to the magnetic field strength. Although we obtained an excellent fit to the \ion{Ca}{ii} 8542 \AA~line in this column, the absolyte temperature residual in i3 is approximately $\Theta\sim$114 K on average between $\log \tau\sim[-4.8,-3.0]$ which is larger than the residuals in i4 ($\Theta\sim80$ K), or even i1 and i5 which are accurate to ($\Theta\sim30-40$ K). Adding the mm points in i4 greatly improved the inversion compared to i3 from $\log \tau=-4.8$ and it is in fact the scheme that performed the best at the nodes $\log\tau=-5.2$ and $\log\tau=-5.6$. At these locations the schemes that performed the worst in the chromosphere were i2 and i6 with an average error of $\Theta\sim$1750 K between $\log \tau\sim[-6.0,-5.0]$ despite the mm-continuum being well fitted. 

At the weakly-magnetic location ($\lvert B_{\rm max}\rvert\sim128$ G) given by the blue square, a 150 K bump above the average value in $T_{\rm model}$ at $\log \tau\sim-4.8$ originates two emission peaks on both sides of the absorption core of the K line. The \ion{Mg}{ii} spectra are best fitted using the i1 and i5 schemes but they are not the ones that best recover $T_{\rm model}$ across the atmosphere. In fact, i6 is the scheme that performs the best overall (at least up to $\log \tau\sim-6.4$) with an average error of $\Theta\sim$80 K between $\log \tau=[-5.0,-4.0]$ and $\sim$250 K at $\log \tau=-5.4$, but it fails to reproduce the temperature peak at $\log \tau=-5.6$ which is underestimated by $\sim1630$ K. The scheme i1 was the one that best recovered this feature with an error of $\Theta\sim$650 K but it shows a much larger error $\Theta\sim$430 K between $\log \tau=[-5.0,-4.0]$.
In this example, i2 clearly provides a more accurate temperature at $\log \tau=-5.2$ than i1 having reduced the discrepancy from $\sim$500 K (in i1) to $\sim$300 K, which is also approximately the same residual level of i6.
In this case the \ion{Ca}{ii} 8542 \AA~line is also very well fitted both with i3-6 but the addition of mm-points and/or extra lines is necessary to improve the accuracy in the layers above $\log \tau\sim-4.8$. We see that in i4 the temperature between $\log \tau=[-6.0,-5.0]$ is much better constrained, but the addition of the mm-continuum made it overestimate the temperatures in essentially all the upper photosphere between $\log \tau=[-4.2,-3.0]$.

Finally, at the dark inter-granule location marked by the green cross we obtained excellent fits to all the spectral points with all the schemes except i2 which was not able to properly fit the continuum between the h and k peaks. Despite that, the accuracy of the inversions in the chromosphere is very different across the schemes with i1, i3 and i5 performing worse in general than their counterparts with added mm-points. At $\log \tau=-5.2$, for example, i1 has a residual of $\Theta\sim500$ K, whereas i2 has a residual of $\Theta\sim$300 K, and i5 has a residual of $\Theta\sim900$ K whereas i6 has a residual of $\Theta\sim400$ K. This emphasizes that a good fit to a single or even several spectral lines does not guarantee a high-fidelity reconstruction of the corresponding chromosphere due to the high degeneracy of the problem. Increasing the wavelength coverage by adding the mm-continuum did help retrieving a more accurate representation of the real temperature profile, but there is still a systematic underestimation of a few hundred kelvin. We should note that, not only at this particular location but also in general, the scheme i4 using only \ion{Ca}{ii} 8542 \AA~ and ALMA to infer the properties of the chromosphere provides a temperature estimate that is of the same order of accuracy of i6 which makes use of a lot more lines. However, the larger combination of lines in i5/i6 provides additional constraints on $v_{\rm los}$ (see appendix Fig. \ref{fig:velocities}), so overall we consider the latter schemes more powerful.

\subsection{Response functions}
\label{sec:RF:BIF}

\begin{figure*}[t] 
\includegraphics[width=\linewidth]{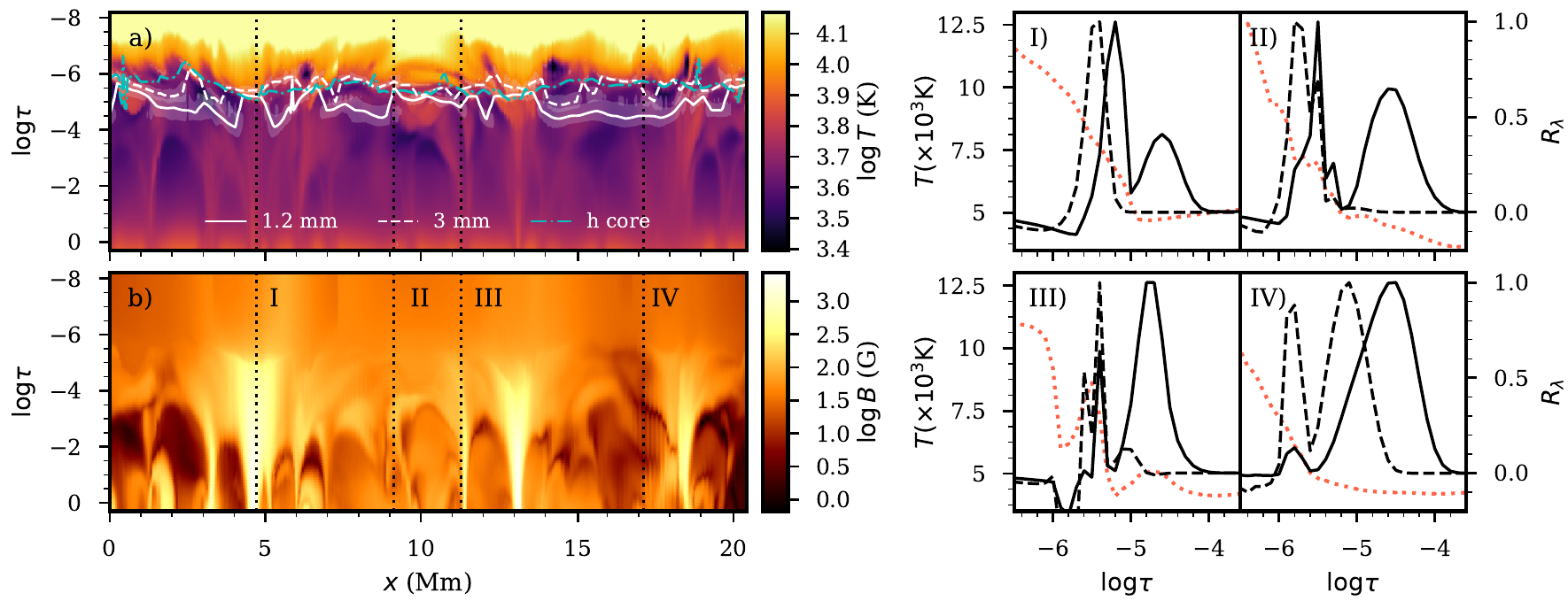} 
\caption{Temperature, response functions to temperature at mm-wavelengths and magnetic field strength in a 2D-cut of a Bifrost simulation. \textbf{a)} Temperature (clipped at 15\,000 K) as a function of optical depth; the solid and dashed lines are the maximum response in each column at 1.2 mm and 3 mm respectively (smoothed for visibility), and the filled contours enclose the $R_\lambda$ at half-peak; the peak response the core of \ion{Mg}{ii} h is shown for comparison (dash-dotted) \textbf{b)} Magnetic field strength for context; \textbf{I-IV)} Normalized response functions of the 1.2 mm- (solid) and 3 mm- (dashed) continuum and temperature profile (dotted line) at the selected columns in panels a/b). } \label{fig:ALMA_response}
\end{figure*}

Figure \ref{fig:3rfs} shows the response functions to temperature perturbations of selected frequencies at the locations marked in Fig. \ref{fig:BIF_mosaic}.
The plots show that $R_\lambda$ is stronger in the line wings but maximal at higher optical depths, and weaker in the cores where it reaches lower optical depths. 
We also see that only a few wavelength points close to the line core are sensitive to the temperature variations in the chromosphere. In addition, none of the wavelength points is specially sensitive in the narrow temperature minimum/steep rise layers usually around $\log \tau\sim[-4.8,-4.0]$. This is why we only obtained an accurate $T_{\rm inv}$ at these depths when we fitted multiple lines (i5 and i6) simultaneously, which together have more constraining power in this region. 

For example, at the location marked by the blue square, the h and k lines of \ion{Mg}{ii} are the ones whose $R_\lambda$ is sensitive to the widest range of optical depths from $\log \tau\sim-2$ to approximately $\log \tau\sim-6.0$, whereas the core of \ion{Ca}{ii} 8542 \AA~is sensitive to lower layers with a maximum response up to $\log \tau\sim-5.2$ (similar to \ion{Ca}{ii} H and K, not displayed). The response of the 1.2 mm-continuum has an absolute maximum at $\log \tau\sim-5.3$ and a secondary peak at $\log \tau\sim-4.9$, but the response of the 3 mm band is single-peaked at $\log \tau\sim-5.6$. The response of the mm bands is practically contained within $\log \tau\sim[-6,-4]$, which encloses the range where we found accuracy improvements with the addition of those wavelengths.

In Fig. \ref{fig:BIF_specs} we showed that none of the schemes was able to recover the steady temperature rise in the higher chromosphere above $\log\tau\sim-5.4$ at the red diamond location, even though the spectra are very well fitted. This can be explained by the response functions at this particular location which occur deeper in the atmosphere than the average. The response of the \ion{Mg}{ii} and \ion{Ca}{ii} lines for $\log\tau\la-5.0$ is negligible, but the 3 mm continuum has its peak at $\log \tau=-5.4$. Therefore, $T_{\rm inv}$ is poorly constrained beyond $\log \tau\sim-4.8$ in the i3 scheme since the response to perturbations in higher layers is practically zero, and only a few points close to the very line core convey information of the layers close to $\log \tau\sim-5.0$. The inclusion of the mm-continuum improved the inverted temperature just until $\log \tau=-5.4$ above which the constraints are very weak.

The diverse density/temperature structure in the simulation generates corrugated surfaces with overlapping distributions and significant responses at distinct heights. 
To further illustrate this point, we computed $R_\lambda$ along a diagonal cut running through the two patches of opposite polarity. 

Figure \ref{fig:ALMA_response} shows the temperature stratification along the 2D cut. On top of the temperature map we overlay the maximum and the half-maximum level of $R_{\rm1.2\,mm}$ and $R_{\rm 3\,mm}$. 
In the bottom panel we show the magnetic field strength for context. We see that the optical depth at which $R_{\rm 3\,mm}$ is maximal is always lower than the maximal $R_{\rm1.2\,mm}$, but we also see that these surfaces are highly corrugated depending very much on the local atmosphere from column to column. For example, the maximum $R_{\rm1.2\,mm}$ spans two units of optical depth and has two peaks at $\log \tau \sim -4.6$ and $\log \tau \sim -5.2$. By fitting a Gaussian mixture model to the corresponding distributions of geometrical heights we found a main peak at $z_1\sim700\pm80$ km and a secondary one at $z_2\sim1200\pm270$ km, with 64\% of the maximum $R_{\rm 1.2\,mm}$ lying below 1000 km. The maximum $R_{\rm 3\,mm}$ does not so evidently show a double peak in optical depth scale, but it does so in $z$-scale with peaks at $780\pm84$ km and $1380\pm186$ km, although in this case for 79\% of the columns $R_{\rm 3\,mm}$ peaks at $z>1000$ km. 

The bimodality at 1.2 mm seems to be correlated with the underlying magnetic field since $R_{\rm1.2\,mm}$ tends to peak at lower optical depths in columns with strong kilogauss magnetic fields which can be found, for example, around $x\sim4.5$ Mm and $x\sim13$ Mm, but not always. We also see that over the magnetic field concentrations $R_{\rm 1.2\,mm}$ and $R_{\rm 3\,mm}$ are more localized in height, whereas in more quiet areas (e.g. $x\sim$14-18 Mm) the distributions are broader. These aspects are illustrated on the right panels in Fig. \ref{fig:ALMA_response} that show how the response varies depending on the location along the slice.

\begin{figure}[t]
\centering
\includegraphics[width=\linewidth]{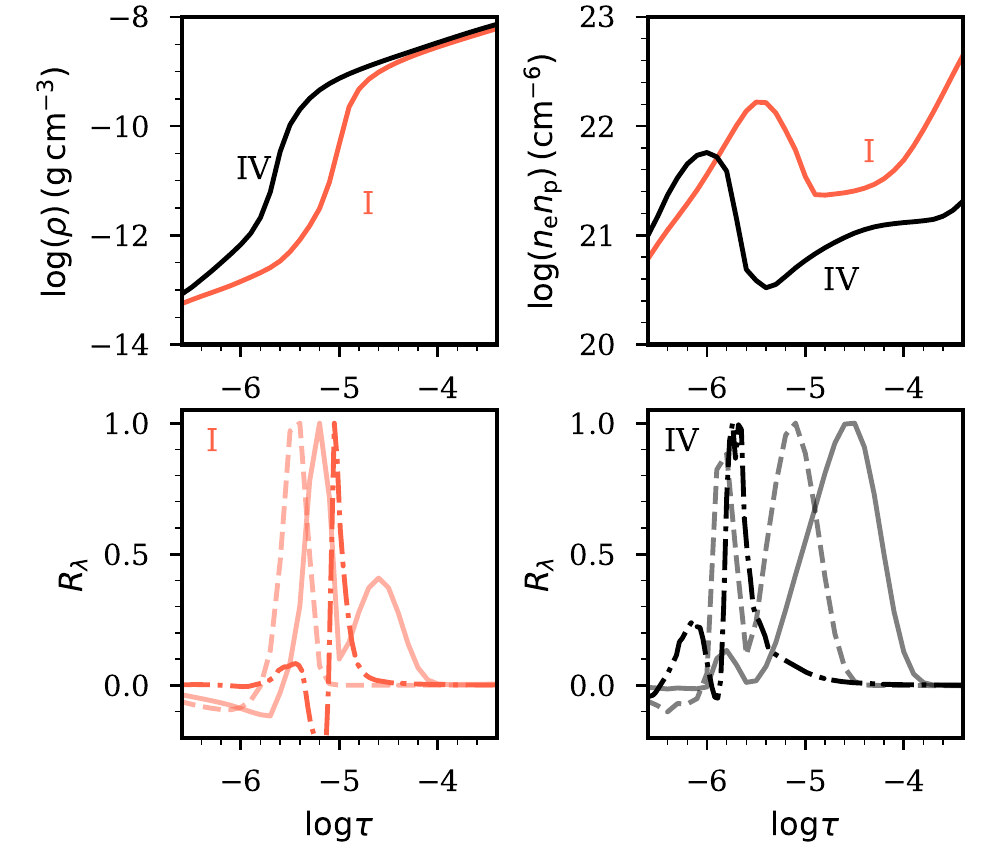}
\caption{Mass density, electron-proton number density and response functions to temperature of the mm-continuum and \ion{Mg}{ii} h. Top: mass-density and electron-proton number density at the magnetic (I) and non-magnetic locations (IV) marked in Fig. \ref{fig:ALMA_response}. Bottom: normalized response functions of the 1.2 mm (solid) and 3 mm-continuum (dashed) and \ion{Mg}{ii} h  core (dashed-dot) at location I (left) and IV (right). } \label{fig:Mg_vs_ALMA}
\end{figure}

Figure \ref{fig:Mg_vs_ALMA} compares mass density and electron-proton number density at two locations marked in Fig. \ref{fig:ALMA_response}, along with the respective responses to temperature perturbations of the 1.2 mm and 3 mm continuum and the core of \ion{Mg}{ii} h. These UV and radio frequencies respond differently according to changes in the opacity.
The chromosphere above strong kG-magnetic elements is evacuated, i.e. has lower mass density than more quiet locations which makes the \ion{Mg}{ii} h and k lines respond to perturbations in deeper layers than the average. On the contrary, the mm-continuum, responds to changes in relatively higher layers because the dominant source of opacity $\chi_{\rm ep} \propto n_{\rm e}n_{\rm p}$, where $n_{\rm e}$ and $n_{\rm p}$ are the electron and proton densities, respectively, is greatly increased compared to more quiet locations. In weakly magnetic locations the difference between the sensitivity of the h core and the mm-continuum is much larger, with the former responding to changes in higher layers at $\log \tau\ga-5.6$, whereas the 1.2 mm, for example, has maximum response at approximately $\log \tau\sim-4.6$ at the selected location. 

There are significant differences between the layers contributing to the response at the core of the h line and the mm-continuum which accentuate in areas of lower magnetic field strength. Above kG-magnetic-field concentrations the response functions of the 1.2 mm and 3 mm continua generally peak at hotter ($T\sim6500-8000$ K) layers with higher electron densities than in more quiet areas (see Fig. \ref{fig:mm_properties} in the appendix).

Figure \ref{fig:BIF_responses} shows the distributions of maximum response functions to temperature perturbations for the selected cube from the Bifrost simulation. In the mm-continuum this corresponds to selected mm-wavelengths whereas for the lines it refers to some wavelength at the line center where $R_{\lambda}$ occurs at the lowest optical depth, i.e., Doppler effects have been taken into account. Apart from the ALMA bands used in the inversions (1.2 and 3 mm) we also computed for completeness $R_\lambda$ at 4.5 mm and 8.5 mm that would correspond to future ALMA bands 1 and 2 under development at the moment of writing. These would allow the study of the upper chromosphere and transition region according to, e.g.,  \citet{2015A&A...575A..15L,2016SSRv..200....1W}. We found that the distributions of maximum $R_\lambda$ move to higher heights with increasing wavelength as expected, although there is considerable overlap between wavelengths.  

In the selected portion of the Bifrost simulation, $R_{\rm 1.2\,mm}$ and $R_{\rm 3\,mm}$ peak at approximately the same range of optical depths as the cores of the \ion{Ca}{ii} and \ion{Mg}{ii} lines with 95\% of the maximum $R_{\rm 1.2\,mm}$ lying between $\log \tau=[-5.8, -4.5]$, while the maximum $R_{\rm 3\,mm}$ lies between $\log \tau=[-6.0, -5.1]$. 
The 1.2 mm continuum responds to temperature perturbations higher up than the core of \ion{Ca}{ii} 8542 \AA~ (and the H and K lines) in about 46\% of the pixels, whereas for the 3 mm continuum that amounts to 82\% of the pixels. The core of \ion{Mg}{ii} lines forms slightly higher than that of the \ion{Ca}{ii} lines and generally above the 1.2 mm continuum since the latter responds to temperature perturbations above the h (and k) line in only 30\% (and 20\%) of the pixels. In 58\% and 49\% of the pixels the 3 mm continuum responds higher than the h and k line respectively. In the majority of the pixels (94\%) the 1.2 mm continuum responds to temperature changes higher up than the UV triplet between the h and k lines.

These results show that these ALMA bands are sensitive to temperature changes from just above the temperature minimum up to the middle-high chromosphere. The range of optical depths $\log \tau\sim[-6.0, -4.5]$ is where we found the most significant improvements with the addition of the mm-continuum to the inversions not surprisingly. It is clear now why $T_{\rm inv}$ substantially deviates from $T_{\rm model}$ from optical depths below $\log \tau=-6.0$ since temperature perturbations in the top of the chromosphere have no impact on the mm-spectra, although that is not the case for the \ion{Mg}{ii} h and k spectra whose $R_{\lambda}$ has a longer tail to even lower optical depths. However, \ion{Mg}{ii} h and k (and \ion{Ca}{ii} H and K) already show a strong scattering decline at these depths meaning that the core of those lines carries little information on the local plasma temperature.

\subsection{The degeneracy between temperature and micro-turbulence}
\label{Section:vturb}

In multi-atom inversions it is, in principle, possible to separately determine both the thermal and microturbulent velocity components because all elements share the same turbulent broadening but undergo different (mass-dependent) thermal broadening. In our inversion schemes $T$ and $v_{\rm turb}$ are fitted simultaneously. 

\begin{figure}[t]
\centering
\includegraphics[width=\linewidth]{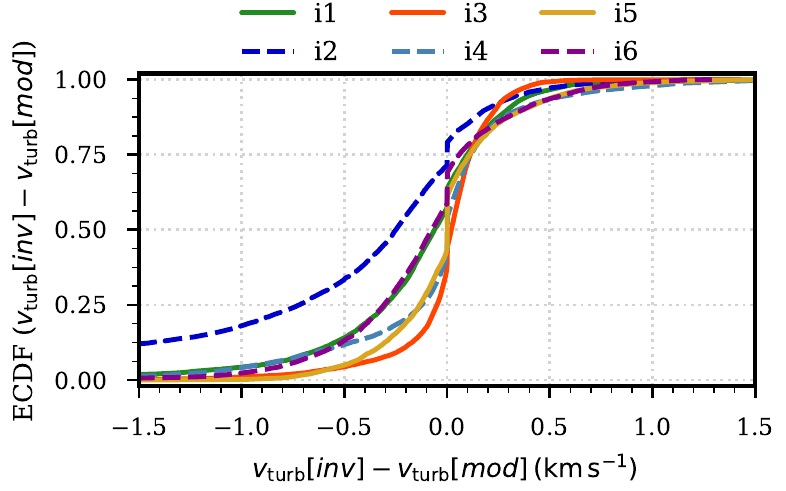}
\caption{Empirical cumulative distribution functions of the micro-turbulence velocity residuals for different inversion schemes.} \label{fig:ecdfs_vturb}
\end{figure}

Figure \ref{fig:ecdfs_vturb} shows the empirical cumulative distributions functions (ECDFs) of micro-turbulence velocity residuals for the different inversion schemes. We found that $v_{\rm turb}$ is more often underestimated than overestimated in all schemes, but the constraints significantly depend on the number of the fitted lines of different atoms.
For example, in i1 $v_{\rm turb}$ is underestimated in 56\% of the pixels, and about 83\% of pixels have a misfit smaller than $0.5 ~\rm km~s^{-1}$. In contrast, in i5 $v_{\rm turb}$ is underestimated in 43 \% and overestimated in 38\%, meaning that 19\% of the pixels have a residual identically zero. In i5, 89\% of the pixels have a misfit smaller than $0.5 ~\rm km~s^{-1}$. 

This is interesting because the temperatures are also more often underestimated (see Fig. \ref{fig:residuals_vl}) in the chromosphere between $\log \tau\sim[-6.0,-4.5]$ where the cores of the \ion{Mg}{ii} and \ion{Ca}{ii} lines respond to temperature perturbations, so it seems that the underestimated temperatures in these layers are not due to overestimated $v_{\rm turb}$. Furthermore, since the absolute residual is mostly smaller than $\sim0.5 ~\rm km~s^{-1}$, this parameter is not expected to have a large impact on the temperature accuracy.

However, we see that schemes fitting the mm-continua (i2, i4 and i6) typically result in larger underestimation of $v_{\rm turb}$ than their counterparts without them, despite improving the inverted temperatures between $\log \tau\sim[-6.0,-4.5]$. 

To further investigate whether our temperature accuracies were limited by the micro-turbulence degeneracy, we synthesized the same spectra setting $v_{\rm turb}=0$, repeated the inversions for the schemes i5 and i6 with exactly the same setup, and compared to the results presented in Section \ref{sec:BIF_results}.

Figure \ref{fig:i5_nv} in the appendix shows the temperature residuals of the inversions for the schemes i5 and i6 with $v_{\rm turb}=0$ at selected optical depth nodes. We found that the temperature residuals are essentially the same in the two inversion modes, i.e., having a constant $v_{\rm turb}$ as an extra free parameter does not greatly impact the determination of temperature. This is because $v_{\rm turb}$ itself is well constrained (better than $0.5 ~\rm km~s^{-1}$) given the number of lines of different atoms that we simultaneously fitted in i5/i6, although this could not be tested for the other schemes. In what regards i5 vs i6, we still found that the mm-continuum provides more accurate chromospheric temperatures when there is no micro-turbulence in the forward calculations. 

\section{Discussion}
\label{Section:Discussion}

\begin{figure}[t]
\centering
\includegraphics[width=\linewidth]{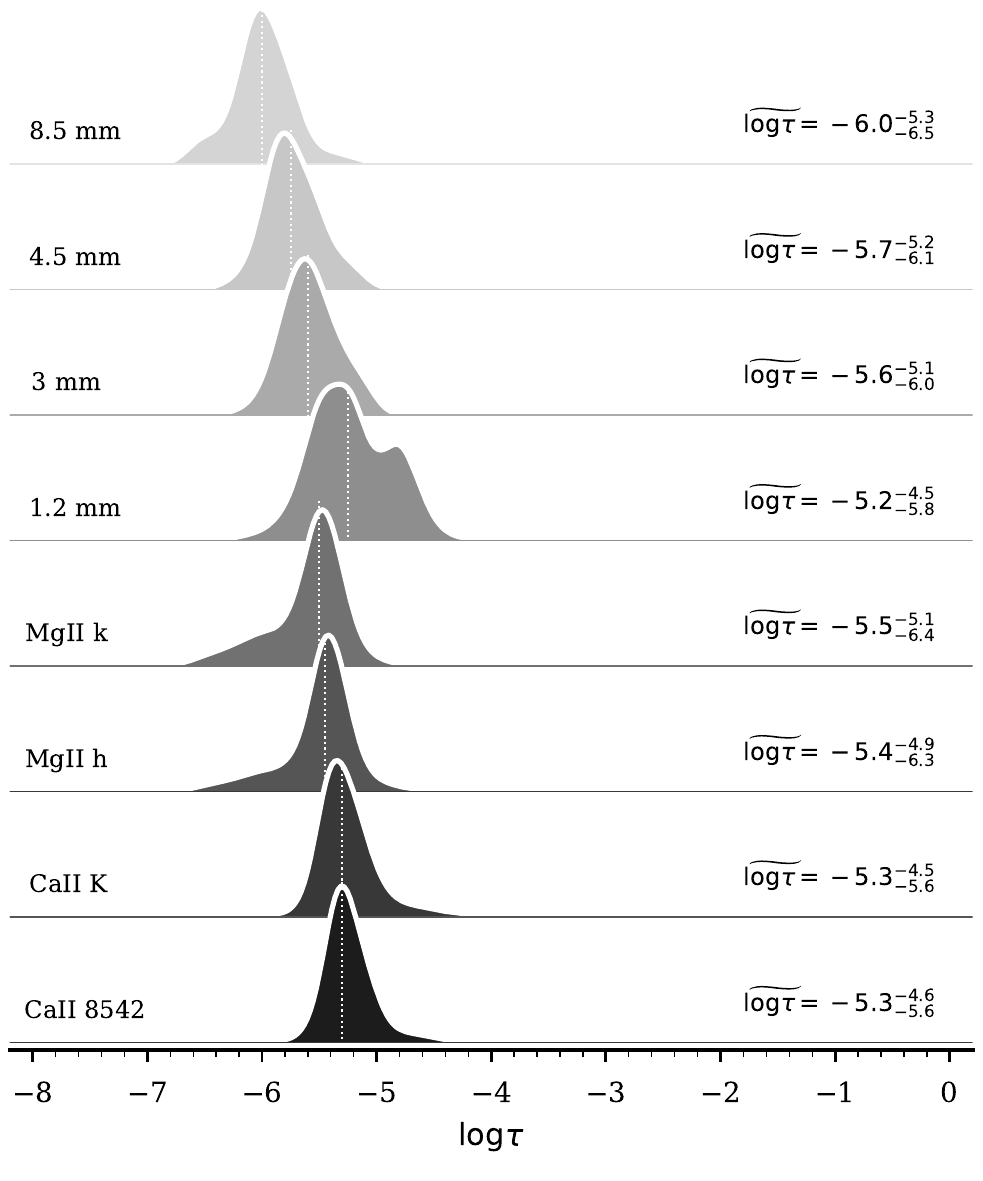}
\caption{Distributions of the maximum response functions to temperature perturbations at selected wavelengths. For the spectral lines we show the distributions at the wavelength where $R_\lambda$ attains the greatest height in each pixel. The values shown correspond to the median, the $2.5^{\rm th}$ and the $97.5^{\rm th}$ percentiles.} \label{fig:BIF_responses}
\end{figure}

\subsection{Temperature constraints from multi-wavelength inversions}

We have used a patch of a Bifrost simulation snapshot to investigate whether the addition of mm-spectra to current inversion schemes improves the reconstruction of the temperature stratification upon solely fitting the profiles of some visible and UV spectral lines. 

We found that mm-wavelength bands can potentially be used to constrain chromospheric temperatures in combination with other diagnostics. This is especially noticeable when comparing the schemes i3 vs i4 which work as "control schemes" to test whether the mm-continuum has any impact on the inversions since, has shown in Fig. \ref{fig:BIF_mosaic}, the higher layers of the atmosphere are completely unconstrained if one uses only \ion{Ca}{ii} 8542 \AA. Although we were able to accurately reproduce such line profiles with i3, the addition of mm-wavelength points in i4 allows to invert the temperature structure in the chromosphere above $\log \tau=-5.0$ where \ion{Ca}{ii} 8542 \AA~loses sensitivity.

The mm-continuum also allows a more accurate inversion of the \ion{Mg}{ii} lines. The 1.2 mm band is especially helpful in setting the gradient between the temperature minimum and the chromospheric plateau, while the 3 mm continuum, which is formed slightly above, provides stronger constraints at higher layers of the chromosphere where the h and k line-cores are only weakly coupled to the local conditions. 
This happens because both 1.2 mm and 3 mm emission have stronger responses to temperature changes at approximately the same heights as the cores of the \ion{Ca}{ii} and \ion{Mg}{ii} doublet lines in a relatively narrow layer of the chromosphere, but the source function of the mm-continuum can be described in LTE by the Planck function ($B_\lambda$) which means we are able to retrieve more information of the local conditions in the higher layers of the chromosphere. 

Our results show that the combination of IRIS and ALMA observations is a powerful tool to diagnose plasma temperatures practically across the entire atmosphere up to the upper chromosphere, specially if one includes photospheric lines in the IRIS spectral window. Inversion codes should make use of mm-data to improve the accuracy of the inverted chromospheric temperatures from \ion{Mg}{ii} lines, while revealing more structure otherwise hidden among the inversion noise. This is of great interest to observers working with ALMA since IRIS usually guarantees observing support.

Interestingly, we also found that a combination of a greater number of UV+optical spectral lines allows inferring reasonably accurate (average error $\sim$30-300 K) temperatures up to the mid-chromosphere ($\log\tau\sim-5.2$) where the the \ion{Ca}{ii} line cores attain maximum response. This emphasizes the importance of the \ion{Ca}{ii} H and K observations with the new SST/CHROMIS instrument \citep{2017psio.confE..85S} whose potential was first shown in \citet{2017ApJ...851L...6R}, since these lines not only greatly improved the inversion of temperatures, but also vertical velocities in the low-mid- chromosphere which were not discussed here (see supplementary Fig. \ref{fig:velocities}). However, this implies that schemes simultaneously fitting a large number of chromospheric lines are not greatly improved by the addition of the 1.2 mm continuum which is sensitive to temperature perturbations at approximately the same heights. The fact that the i6 maps show more structure with higher spatial correlation with respect to the model at $\log\tau\sim-5.6$ is mostly due to the 3 mm continuum whose response peaks around such optical depth.

We found that the response functions to temperature at mm-wavelengths change spatially depending on the underlying temperature/density structure and on the magnetic field to some extent. 
$R_{\rm 1.2 mm}$ spans a larger range of optical depths with two clear peaks at $\log\tau=-4.8$ and $\log\tau=-5.2$, which is lower down in the atmosphere than the typical maximum response of the core of the \ion{Mg}{ii} lines ($\log\tau\sim-5.5$). We also found an offset in the optical depth of the maximum response of the h (and k) core and the mm-continuum depending on the underlying magnetic field strength. 

This behavior is consistent with the effective formation heights defined as the centroid of the contribution functions of the mm-continuum shown in \citet{2015A&A...575A..15L} using the same enhanced network model. The authors found that on average the contribution maximum at $\lambda=1$ mm and $\lambda=3$ mm occurs at 900 km and 1500 km respectively, which are slightly higher than the values reported in \citet{2007A&A...471..977W} who found 730 km and 960 km, respectively, from another simulation.
The resemblance between contribution functions and response functions to temperature disturbances happens because temperature is the dominant parameter in mm-continuum formation.  
We note that collapsing the entire contribution function to a single value hides the fact that there are in fact multiple heights contributing the the mm-radiation with some overlap between wavelengths. This means that in a quiet-atmosphere ALMA bands provide constraining power not at a single well-defined height in the chromosphere but at a wide range of heights anywhere between just above the temperature minimum and the upper chromosphere depending on the opacity. 

In the Bifrost snapshot there are systematic differences between the responses of the \ion{Mg}{ii} line-cores and the 1.2 mm radiation with the latter being more influenced by perturbations in lower layers of the chromosphere than the former in the majority of the pixels which correspond to quiet atmospheres. On top of strong magnetic elements that trend can reverse, with the line cores responding to perturbations deeper in the atmosphere as consequence of opacity changes. 

On top of kG- magnetic field concentrations the mm-continuum responds to temperature perturbations at lower optical depths in tendentially hotter and electron-denser layers of the chromosphere. 

In principle, including the 4.5 mm and 8.5 mm bands (not available yet for ALMA observations) could help constraining even higher layers of the atmosphere since $R_\lambda$ peaks at higher heights at those wavelengths, although we have not performed inversion experiments with those. Similarly, including bands 4 and 5 sampling between 1.4-2.4 mm (not available for solar observations) could also offer additional temperature constraints, while allowing the so-called tomographic or "volume imaging" of the chromosphere in the words of \citet{2007A&A...471..977W}.

\begin{figure*}[t]
\centering
\sidecaption
\includegraphics[width=120mm]{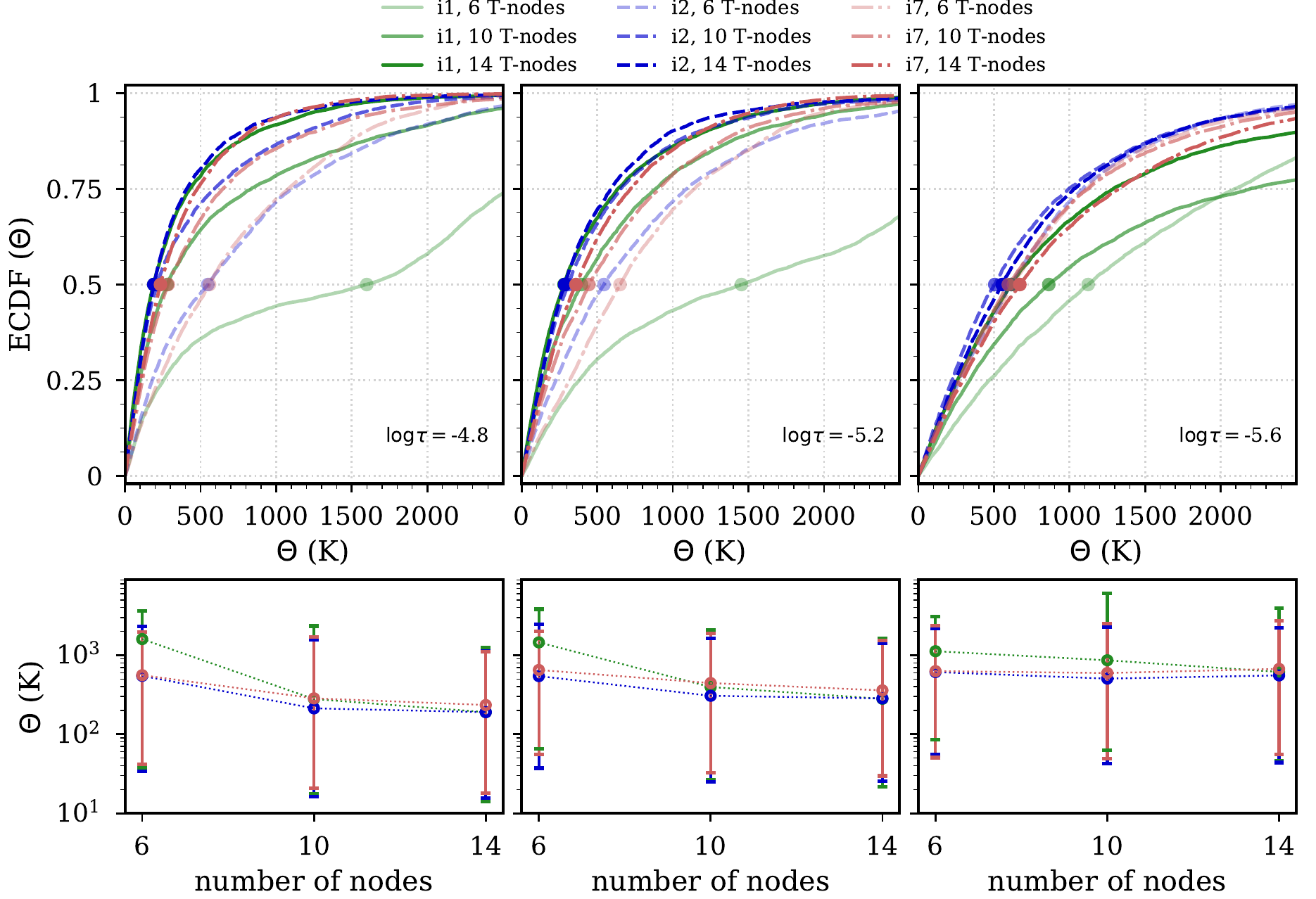}
\caption{Temperature residuals versus number of nodes for different inversion schemes. Top row: empirical cumulative distribution functions of the absolute residuals (Eq. \ref{eq:error2}) of the schemes i1 (solid green), i2 (dashed blue) and i7 (dashed-dot red) for different number of temperature nodes at three selected optical depths in the chromosphere; Bottom row: corresponding median (circle markers) and the interval (vertical lines) containing 90\% of the distribution of absolute residuals. } \label{fig:nodes}
\end{figure*}

\subsection{Caveats}
\label{Section:limitations}

We note that the estimated uncertainties are model dependent. The actual accuracy gain in real observations will depend on a number of factors and it is hard to quantify. These include slight temporal/spatial mismatches between different observatories, the signal-to-noise ratio, the absolute flux calibration uncertainties and the target on the solar surface.  

In fact, strictly speaking these results refer to a enhanced-network-like region of the Sun so it is unclear whether the same results would be found in less quiet regions. It could be that in active regions and plage the hotter and denser plasma ensures a better coupling of the source functions to temperature, thus allowing better inversions. On the other hand, it might not be possible to ignore synchrotron contribution at mm-wavelengths at certain locations in active regions.

On the technical side we should mention that it is not currently possible to perform simultaneous ALMA observations at two distinct bands as we have assumed, since there are calibration overheads that typical amount to at least 30 min between bands at the moment of writting, although it is not clear whether this could change in the future. Therefore, it might not be possible to benefit from the information of both bands on the same target at the same time. This is definitely a problem for the most dynamic, short-lived phenomena in the chromosphere such as UV-bursts and flares occurring on much smaller time scales of a few minutes. However, in some cases it may be useful to perform some sort of spatial averaging to simply measure the average brightness temperature over larger scales. 

In fact, some solar features have enough structural stability over longer periods of time which may justify the usage of such time sparse data. For example, as in the sunspot analysis of \citet{2017ApJ...850...35L} who combined ALMA observations with both bands 3 and 6 taken in two different days to constrain the average properties of a sunspot's umbra and penumbra while comparing to model predictions.

Nonetheless, we performed an addition inversion test (i7) with IRIS+ALMA synthetic data, this time using only the continuum around 1.2 mm (band 6). The inverted maps are shown in the supplementary Fig. \ref{fig:iris_alma12}. We found that even if we include only this band, we still obtain clear improvements on the inverted temperatures in the low-mid- chromosphere. This test could have been done using only band 3 with a similar outcome, although they actually respond to temperature perturbations at slightly different layers (see Fig. \ref{fig:BIF_responses}). Band 6 was preferred since it has a higher spatial resolution, currently at 0.63\arcsec for a maximum baseline of 500 m (at the moment of writing), which is closer to the IRIS NUV resolution of approximately 0.33\arcsec, therefore it can more realistically provide reliable results on real data at the current ALMA capabilities. These are expected to improve in the future.

The node placement was extensively investigated and ultimately ruled out with the inclusion of higher number of nodes densely sampling the chromosphere. We investigated this issue in greater detail with the schemes i1 and i2 as shown in Figure \ref{fig:nodes}. The ECDFs of the absolute temperature residuals ($\Theta$) for different inversion cycles show that the accuracy of the inversions generally increases with the increase of number of nodes, but the bulk of the pixels is not very affected by increasing from ten to fourteen nodes in temperature, so we would not expect to obtain substantially different results with even more nodes since the improvement in the residuals already seems to be damped. For example at $\log \tau=-4.8$ the median of the absolute residuals in i1 decreased by 1300 K from six to 10 nodes, but it only further decreased another 100 K from 10 to 14 nodes. In i2, from 10 to 14 nodes the median accuracy gain was only 15 K, while the changes in the tails of the distributions are insignificant.

Figure \ref{fig:nodes} also shows that inversions with mm-wavelength points are always more accurate than inversions of only \ion{Mg}{ii} lines at the selected optical depths in the chromosphere. In particular, for a low number of nodes the accuracy gain by simply adding ALMA bands is quite striking (1000 K in the median). Progressive inversion cycles improve both i1 and i2, and eventually with a sufficiently high number of nodes the accuracy of these inversion schemes is of the same order in the layers $\log \tau\sim[-5.0,-4.0]$. However, higher up, e.g. at $\log \tau=-5.6$, i2 is still more accurate by some 130 K on average, and the temperature maps are a lot smoother because the tail of the distribution of residuals for the overestimated temperatures is significantly reduced (by $\sim$ 1700 K) as shown in the corresponding bottom panel of Fig \ref{fig:nodes}.

The ECDFs for i7 show that even the addition of just the continuum around 1.2 mm helps the inversions to reduce the errors across the chromosphere. As expected, in the low chromosphere one does not lose a great degree of accuracy by not simultaneously using the 3 mm continuum since it has much weaker response at $\log \tau>-5.0$. For example, at $\log \tau=-4.8$ i2 and i7 provide essentially the same result for 6 nodes in temperature which is significantly better than i1, but for 14 nodes there is no significant difference in the accuracy of the three schemes with i2 having only a minor ($\sim50$ K) improvement on the median error compared to i7. However, higher up in the chromosphere, e.g. at $\log \tau=-5.6$ we do see quite a noticeable difference between i7 and i2 because the 3 mm continuum provides stronger constraints at those optical depths. 
In this case, increasing the number of nodes from 10 to 14 actually caused a larger error in a portion of the pixels in i7 since, for example, in cycle 2 75\% of the pixels had a residual smaller 1100 K whereas in i7 that fraction is 69\%, although this could be attributed to the inversion settings. 

It is clear that the 3 mm continuum is necessary to recover temperatures at even lower optical depths where the response of the 1.2 mm continuum and UV/optical lines is lower. Therefore we speculate that had we chosen to recompute i4 or i6 with only the 1.2 mm-band, we would have arrived to the same conclusion. This test confirms the intuition provided by the responses functions: the 1 mm continuum is useful to set the gradient between the temperature minimum and the low-chromosphere while improving the accuracy of the inverted temperatures at lower layers around $\log \tau\sim-5.0$, whereas the 3 mm continuum acts in the mid-chromosphere since it responds to temperature perturbations slightly above.

\subsection{Conclusions}
\label{Section:Conclusions}

We have used a snapshot of a 3D radiation-MHD simulation with Bifrost to investigate whether ALMA can help the inversion of commonly used non-LTE spectral lines, while essentially allowing a better representation of the real temperature structure of the solar atmosphere. 
Our main findings can be summarized as follows:

\begin{itemize}
\item Inversions of a combination of lines and mm-continuum can retrieve with reasonable accuracy, higher in the photosphere and lower in the chromosphere, the temperature structure of a solar-like atmosphere.
\item The mm-continua helps the inversions of commonly used non-LTE lines namely \ion{Mg}{ii} h and k and \ion{Ca}{ii} H, K and 8542 \AA~to retrieve a more accurate representation of the temperature stratification in the mid-upper chromosphere where those lines are only weakly coupled to local conditions. The addition of mm-wavelengths is very useful to remove outliers from the error distributions and to substantially reduce the "inversion noise", revealing more spatial structure.
\item The 1 mm continuum responds to temperature perturbations typically around $\log \tau\sim-5.0$, and thus is useful to set the gradient between the temperature minimum and the low-chromosphere, whereas the 3 mm continuum has its maximal response slightly above at $\log \tau\sim-5.6$ on average, and extending up to $\log \tau\sim-6.0$, constraining the mid-upper chromosphere.
\item The response of the mm-continua changes spatially depending on the underlying temperature and density structure and to the magnetic field strength to some extent, since we found that on top of kG-magnetic field elements the peak response of the ALMA bands moves to lower optical depths at hotter layers of the chromosphere with higher electron densities than in more quiet areas. This is usually the opposite behavior of the \ion{Mg}{ii} line cores which have deeper responses at such locations where the mass density in the chromosphere is lower.
\item There are systematic differences in the heights contributing the response to temperature perturbations between the \ion{Mg}{ii} lines and the mm-continua. In quiet (weakly magnetic) areas there is a larger gap between the optical depth of the peak response of the h and k cores and the 1.2 mm radiation with the former responding to temperature changes well above the 1.2 continuum, whereas above strong magnetic elements this behavior reverses.
\item The 1.2 mm band is expected to give the most reliable results at the current ALMA spatial resolution capabilities ($0.6$\arcsec), and we have shown that it helps inversions of the \ion{Mg}{ii} lines and even \ion{Ca}{ii} 8542\AA~alone. However, simultaneous inversions of \ion{Mg}{ii} and \ion{Ca}{ii} lines are not as greatly improved by the addition of the 1.2 mm continuum since the latter is sensitive to temperature changes in approximately the same layers as the cores of the H, K and 8542 \AA~ lines in most cases. Provided that enough inversion cycles are performed with a sufficiently high number of nodes, together with regularization to improve convergence, inversions of these lines are able to attain the same order of accuracy up to $\log\tau\sim-5.2$. This emphasizes the importance of the \ion{Ca}{ii} H and K observations with the new SST/CHROMIS instrument since these lines not only greatly improved the inversion of temperatures, but also vertical velocities in the low-mid- chromosphere.
\item Depending on specific science goals it may not be possible to run multiple inversions cycles on large patches for different time steps due to the prohibitive computational costs. In that case, inversions with the mm-continuum provide relatively more accurate temperatures for a smaller computational effort since they achieve good a degree of accuracy with less cycles/nodes. The actual accuracy gain in real observations will depend on a number of factors such as slight temporal/spatial mismatches between different observatories, the signal-to-noise ratio, the absolute flux calibration uncertainties and the target itself.
\item The 3 mm band is needed in any case to constrain even higher layers of the chromosphere where the lines show weaker responses and their source functions are dominated by scattering.
\end{itemize}

The next decade is likely to see improvements in our understanding of the solar chromosphere now that the community has access to the full electromagnetic spectrum of the Sun from the radio to the ultraviolet in high resolution. This will necessarily require coordinated, multi-wavelength campaigns along with inversion codes, such as STiC, to infer the thermodynamical state of the plasma from such observations. 

\begin{acknowledgements}
The computations were performed on resources provided by the Swedish National Infrastructure for Computing (SNIC) at the High Performance Computing Center North (HPC2N) at Umeå University.

This research made use of Astropy, a community-developed core Python package for Astronomy (Astropy Collaboration, 2018). 

JdlCR is supported by grants from the Swedish Research Council (2015-03994), the Swedish National Space Board (128/15) and the Swedish Civil Contingencies Agency (MSB). This project has received funding from the European Research Council (ERC) under the European Union's Horizon 2020 research and innovation programme (SUNMAG, grant agreement 759548).
\end{acknowledgements}

%\bibliographystyle{aa}  

% _____________________________________________________ %
\begin{appendix}
% _____________________________________________________ %
\section{Supplement figures}

\begin{figure*}[t]
\centering
\includegraphics[width=\linewidth]{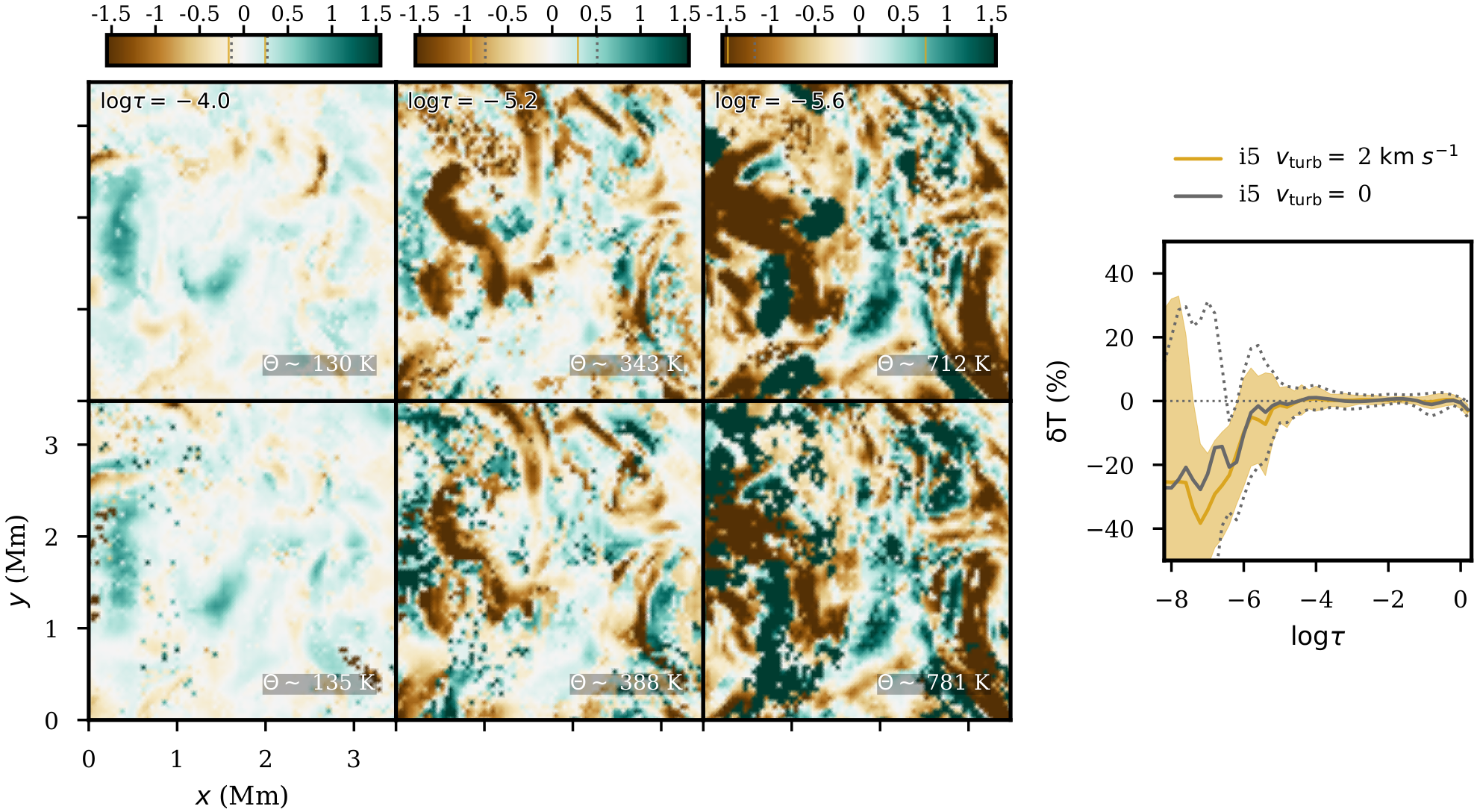} \\
\includegraphics[width=\linewidth]{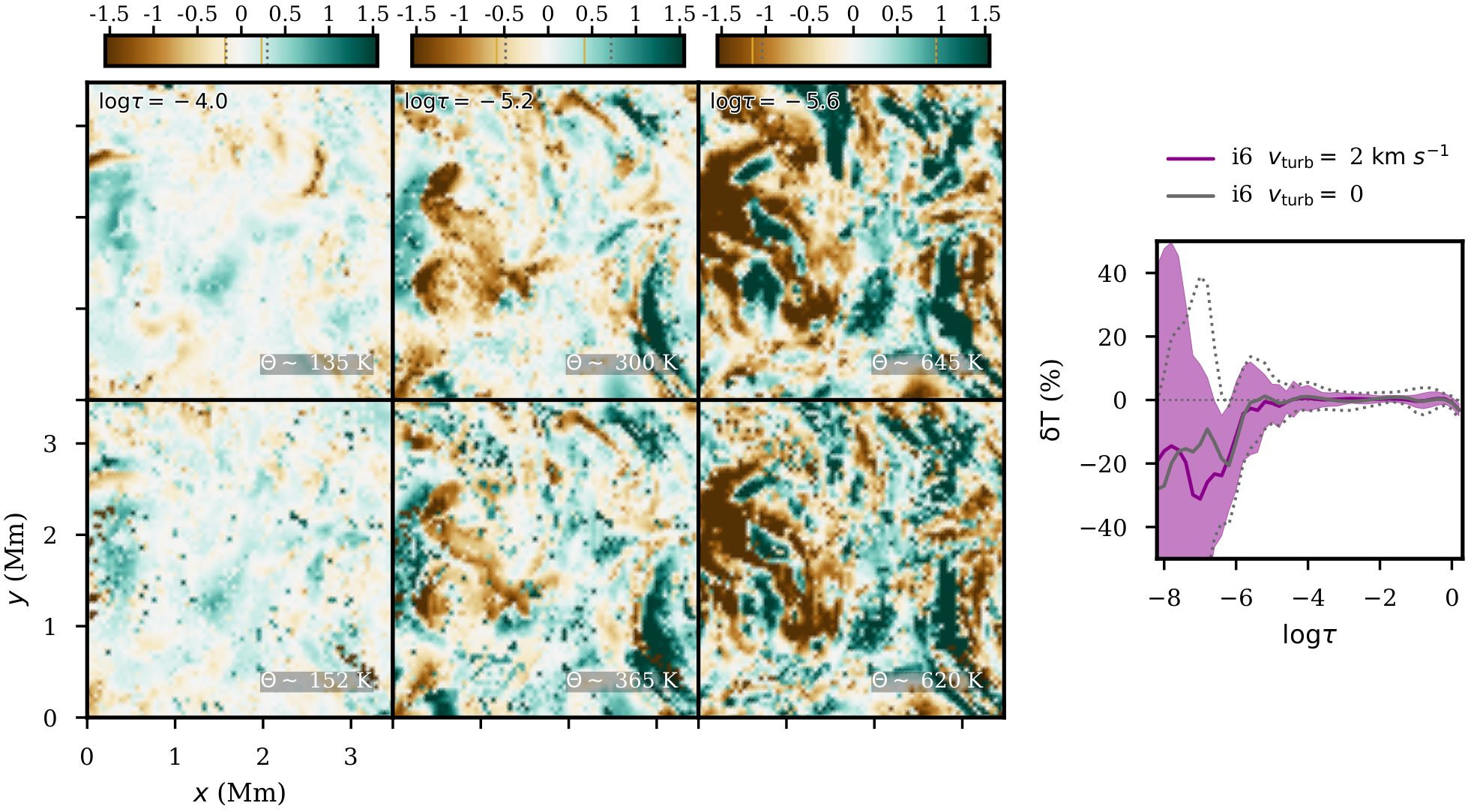}
\caption{Temperature residuals of two inversion schemes at selected optical depths for synthetic spectra with null micro-turbulence. In each group of panels, the top row correspond to the fit to synthetic data with constant $v_{\rm turb}=2~\rm km~s^{-1}$ and the bottom row is fit to synthetic data with $v_{\rm turb}=0$. The colorbars are in units of $10^3$ K. The rightmost panels show the relative error for the two inversion modes as in Fig. \ref{fig:error}.} \label{fig:i5_nv}
\end{figure*}

\begin{figure*}
\centering
\includegraphics[width=\linewidth]{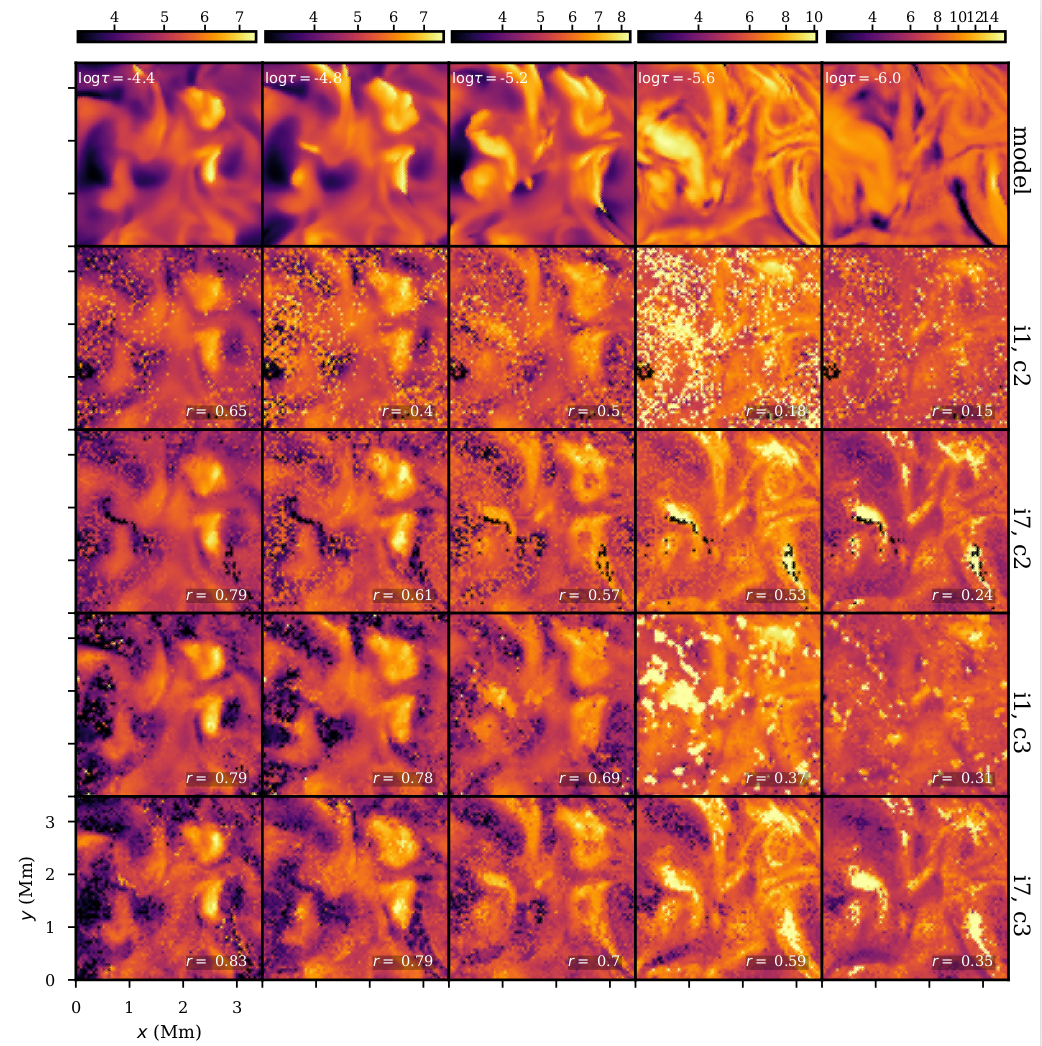}
\caption{Model and inverted temperatures as function of optical depth in a patch of a Bifrost simulation snapshot. The top row shows the temperature in the model at selected optical depths in the chromosphere; we compare the schemes i1 and i7 for cycle 2 (10 temperature nodes) and cycle 3 (14 temperature nodes). The colorbars are in units of $10^3$ K. $r$ is the correlation coefficient. } \label{fig:iris_alma12}
\end{figure*}

\begin{figure*}
\centering
\includegraphics[width=0.98\linewidth]{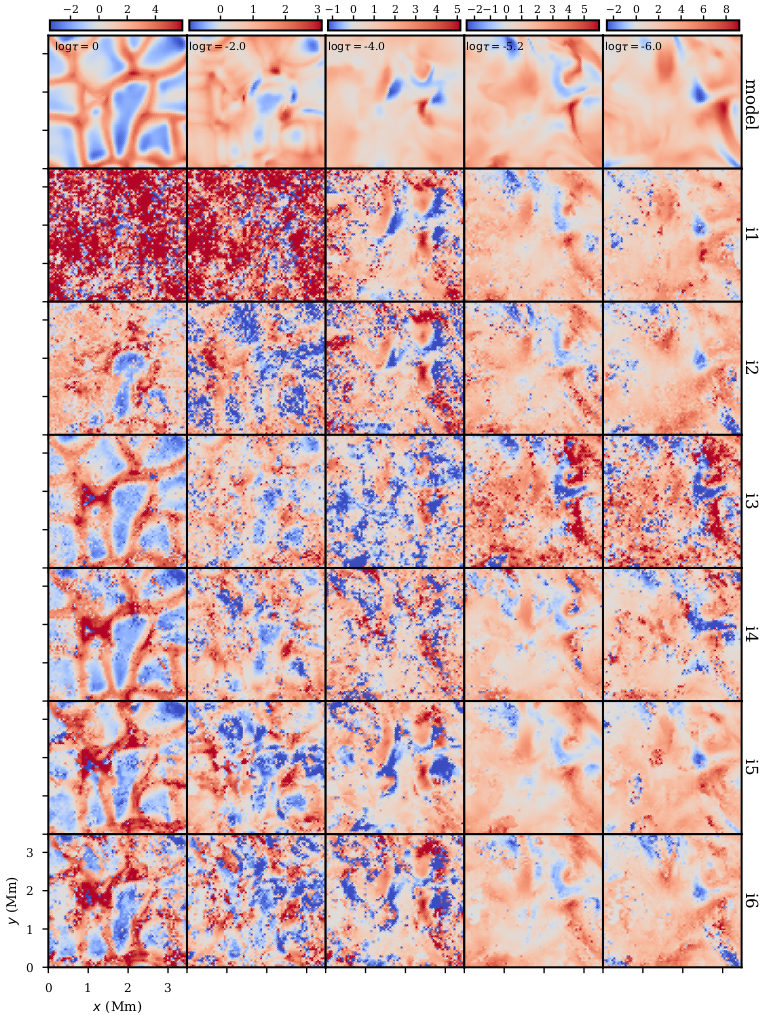} 
\caption{Model and inverted line-of-sight velocity as function of optical depth in a patch of a Bifrost simulation snapshot. The colorbars are in units of $\rm km~s^{-1}$.} \label{fig:velocities}
\end{figure*}

\begin{figure}
\centering
\includegraphics[width=\linewidth]{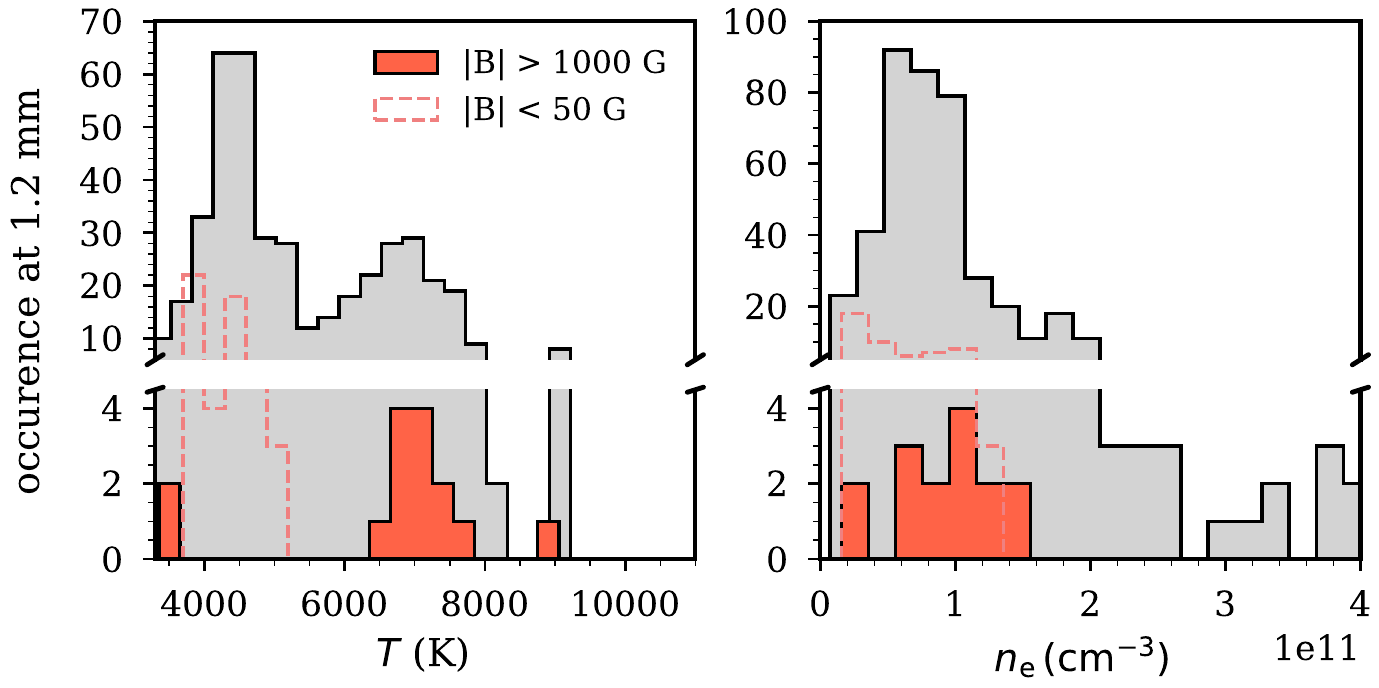}\\
\includegraphics[width=\linewidth]{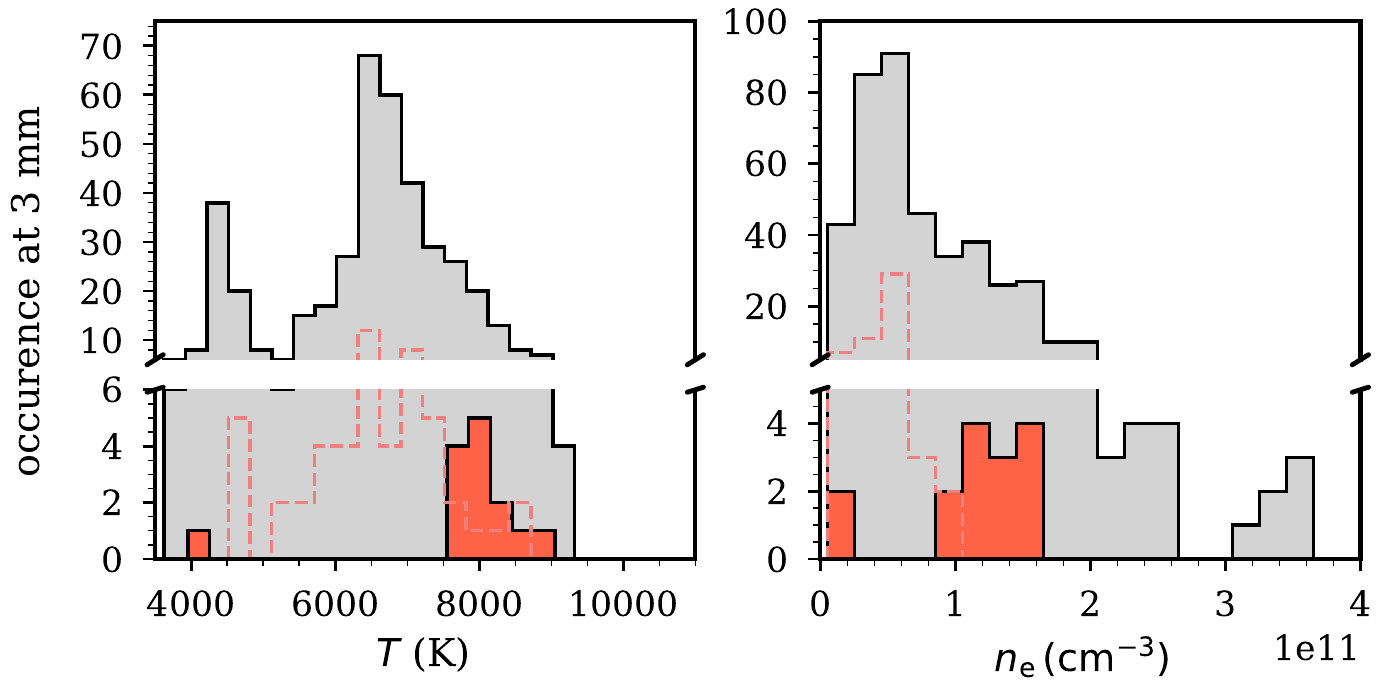}
\caption{Distributions of temperature and electron density at the optical depth of the peak response at two mm-wavelengths. The left panels compare the temperature histograms along the lines drawn in panels a/b) of Fig. \ref{fig:ALMA_response} for all magnetic field strengths (gray-filled), weak-magnetic elements ($|B_{\rm max}| <50$ G, dashed orange) and strong magnetic concentrations ($|B_{\rm max}| >1000$ G, orange-filled). The right panels show an analogous comparison in terms of electron densities. } \label{fig:mm_properties}
\end{figure}

\end{appendix}

\end{document}